\documentclass[journal]{IEEEtran}
\usepackage[utf8]{inputenc}
\usepackage{cite}
\usepackage{graphicx}
\graphicspath{ {Images/} }
\usepackage{amsmath}
\usepackage{mathrsfs,amsthm,amsmath,amssymb}
\usepackage{textcomp}
\usepackage{graphicx,multirow}
\usepackage{balance}
\usepackage{cite}
\usepackage{float}
\usepackage{gensymb}
\usepackage{eso-pic}
\usepackage{algorithm}
\usepackage{algorithmic}
\usepackage{xcolor}
\usepackage{scalerel}
\usepackage{tikz}
\usetikzlibrary{svg.path}

\newcommand{\bs}{\boldsymbol}
\newcommand{\mb}{\mathbf}
\newcommand{\jj}{\boldsymbol{j}}

\newcommand{\revisioncolor}{black}

\newcommand{\inquirycolor}{black}

\newcommand{\revcolor}{black}
\newcommand{\quirycolor}{black}

\DeclareMathOperator{\st}{s.t.}
\DeclareMathOperator{\tr}{Tr}
\DeclareMathOperator{\diag}{diag}
\DeclareMathOperator{\vect}{vect}
\DeclareMathOperator{\re}{Re}
\DeclareMathOperator{\rank}{Rank}

\graphicspath{ {Images/} }
\IEEEoverridecommandlockouts
\begin{document}
\title{IRS-aided Large-Scale MIMO Systems with Passive Constant Envelope Precoding }



\author{Yasaman~Omid,~\IEEEmembership{Student Member,~IEEE}, Seyyed MohammadMahdi~Shahabi,~\IEEEmembership{Student Member,~IEEE}, Cunhua~Pan,~\IEEEmembership{Member,~IEEE}, Yansha~Deng,~\IEEEmembership{Member,~IEEE}, Arumugam~Nallanathan,~\IEEEmembership{Fellow,~IEEE}

 \thanks{Yasaman Omid, Cunhua Pan  and Arumugam Nallanathan  are with the School of Electronic Engineering and Computer Science, Queen Mary University of London, U.K. (e-mail: y.omid@qmul.ac.uk; c.pan@qmul.ac.uk; a.nallanathan@qmul.ac.uk).}

 \thanks{Seyyed MohammadMahdi Shahabi is with the Department of Electrical Engineering, K. N. Toosi University of Technology, Tehran, Iran (e-mail: shahabi@ee.kntu.ac.ir).}

\thanks{Yansha Deng is with the Department of Engineering, Kings College London, U.K. (e-mail: yansha.deng@kcl.ac.uk).}}

\maketitle
\begin{abstract}
In this paper, an intelligent reflecting surface (IRS)-aided large-scale MIMO system is investigated in which constant envelope precoding (CEP) is utilized at each base station (BS). It is both cost-effective and energy-efficient to implement CEP  in a large-scale antenna array. We aim to optimize the discrete phase shifts at both the BS and the IRS to minimize the sum power of multi-user interference (MUI) in the system
via our proposed three algorithms. For the sake of simplicity, a simple single-cell scenario is considered, where the optimization of the BS and IRS phase shifts is solved by a low-complexity trellis-based algorithm. Then, this algorithm is extended to a multi-cell scenario, where the precoding operation in each BS is performed individually. With the aid of stochastic optimization method, a low-overhead trellis-based solution is proposed which has better performance than the first one. Finally, we solve the optimization problem via the semi-definite relaxation (SDR)  scheme, to serve as a performance benchmark for the proposed algorithms. Meanwhile, interference and complexity analysis is provided for the proposed algorithms. Numerical results demonstrate that while the performance of the trellis-based algorithms is negligibly lower than that of the continuous-phase SDR-based solution, the computational complexity and the implementation cost of the former is much lower than the latter, which is appealing for practical applications.
\end{abstract}
\begin {IEEEkeywords}
Large-Scale MIMO, Intelligent Reflecting Surface, Reconfigurable Intelligent Surface, Constant Envelope Precoding, Massive MIMO.
\end{IEEEkeywords}
\section{Introduction}
\label{intro}
\IEEEPARstart{T}{he} grand requirement for data traffic in the next generation of wireless communications necessitates innovative technologies to be developed. To address this issue, several energy-efficient wireless solutions have been presented, among which large-scale multiple input multiple output (MIMO) systems are widely known to be effective in providing high throughput, reliability as well as energy-efficiency \cite{Marzetta2010NoncooperativeAntennas,Rusek2013}. The primary idea of large-scale MIMO systems is to equip the base station (BS) with very large antenna arrays to simultaneously serve multiple user terminals in the same time/frequency resources.

Even though various advantages have been enumerated for the large-scale MIMO systems, their practical implementation faces certain challenges. On one hand, the power amplifiers (PA) used at the BS are expected to be highly power efficient. Due to the trade-off between a PA's efficiency and its linearity, it is preferred that a non-linear highly power efficient PA should be employed. On the other hand, the hardware complexity of such systems is high, since each antenna is connected to a radio frequency  (RF) chain with a power-consuming PA. \textcolor{\revisioncolor}{Exploiting the concept of constant envelope precoding (CEP)\cite{Mohammed2013Per-AntennaSystems}, both these issues could be thoroughly addressed.}  Firstly, as the efficiency of a PA depends upon the amount of required back-off caused by the value of peak-to-average power ratio (PAPR) to compensate for the non-linear distortions, constant envelope (CE) input signals could be utilized to achieve the minimum back-off and thus the maximum efficiency. Secondly, the number of PAs at a BS using CEP is reduced to merely one, as the outgoing amplitude of all signals from all BS antennas is the same.

The CEP structure was initially introduced in \cite{Mohammed2013Per-AntennaSystems} for a single-cell MIMO system. The authors in \cite{Kazemi2017} considered a more practical realization for the single-cell CEP by assuming that the antenna phases are selected from a discrete set of angles through a trellis-based algorithm. This low-complexity method gives comparable performance to that of the non-linear optimization method \textcolor{\inquirycolor}{by the mesh adaptive direct search (NOMAD)}. In \cite{Shahabi2019IET,Shahabi2019PhyComm,Shahabi2018}, the multi-cell scenario for CEP was considered and various algorithms were proposed to minimize the system overhead as well as the power of the interference terms caused by channel estimation error and pilot contamination. In addition, in order to reduce the hardware complexity, a \textcolor{\revisioncolor}{mean square error} (MSE)-based linear hybrid precoding and equalization design is employed in a full duplex (FD) massive MIMO system in \cite{Mai2016JointSystems}, where the number of RF chains is smaller than the number of the BS antennas. \textcolor{\revisioncolor}{Inspired by} this work, the authors in \cite{Omid2018IST} presented a CEP-based FD massive MIMO system in which the precoding design  not only requires minimal number of RF chains, \textcolor{\revisioncolor}{but also maintains a tolerable interference level.}

Another promising technique which has been proposed recently is the intelligent reflecting surface (IRS) which is a low-cost spectrum and energy efficient wireless solution that achieves high performance by reconfiguring the wireless propagation environment \cite{Pan2019}. Specifically, an IRS is a meta-surface with artificial passive radio elements that reflect the RF waves towards a specific direction via passive reflection beamforming. Hence, deploying such a structure only requires a large number of passive phase shifters (PS) without any PAs, which makes it a power efficient technology. \textcolor{\inquirycolor}{In addition, IRSs are cheap and easily integrated into the communication environment such as buildings facades and ceilings, laptop cases, and so on \cite{Huang2018}.} Motivated by these beneficial features, the topic of IRS-aided large-scale MIMO systems has seen a significant surge in popularity among researchers in the field of 5G communications and beyond.

As the IRS has become one of the \textcolor{\revisioncolor}{hottest} topics in wireless communications, various passive beamforming schemes have been studied in the literature for different IRS-aided systems.
\cite{Wu2019IntelligentBeamforming}, a MISO system was considered, in which an IRS is applied to assist the communication from the multi-antenna access point (AP) to multiple single-antenna users. Specifically, \textcolor{\revisioncolor}{a single-user case} was  considered  and  two  solutions for IRS phase optimization problem were  proposed  based  on \textcolor{\revisioncolor}{semi-definite relaxation} (SDR) and alternate optimization techniques.
Then, the single-user scenario was extended to a multi-user scenario.
In  \cite{Pan2019a}, the high beamforming gain brought by IRS was used for simultaneous wireless information and power transfer (SWIPT)-aided system. To maximize the weighted sum-rate, the authors used the block coordinate descend (BCD) algorithm to decouple the main optimization problem into several sub-problems, and they provided low-complexity iterative algorithms with guaranteed convergence to the Karush-Kuhn-Tucker (KKT) point.
The authors in \cite{Zhou2019} studied the the IRS-aided downlink multi-group multicast communication system, where the sum-rate of all the multicasting groups were maximized by joint optimization of BS precoding matrices and the phase shifts of the IRS.
In \cite{Bai2019}, the authors studied the application of IRS in mobile edge computing systems, where the latency was minimized through joint optimization of IRS phase shifts, offloading selections and decoding vectors at the BS.  The BCD technique was used to decouple the main optimization problem into two sub-problems, and then low-complexity iterative algorithms were provided that could achieve KKT-optimal point.
The authors in \cite{Guo2019} considered an IRS-aided MISO downlink communication system, and they aimed to maximize the weighted sum-rate (WSR) of all users by jointly optimizing the active beamforming at the BS and the passive beamforming at the IRS subject to the transmit power constraint at the BS.
The active beamforming at the BS was solved via the fractional programming method and then three different approaches with closed-form expressions were proposed for the IRS passive beamforming.
In \cite{Pan2019}, an IRS was employed for aiding the transmission of cell-edge users in a multiple-cell MIMO system as well as alleviating the inter-cell interference.
In \cite{Wu2018}, the authors considered the joint active and passive discrete beamforming optimization problem.
In \cite{Huang2018a}, the energy efficiency maximization problem was studied through the joint power allocation and discrete phase optimization.
It was shown that the system performance is enhanced immensely in terms of energy efficiency compared to the conventional relay-assisted communications.
Further research on IRS can be found in \cite{Liu2020}, \cite{Wu2019c}, \cite{Zhou2020} and \cite{Zhou2019a}.

It is noteworthy that all the existing papers considered active beamforming at the BSs. Digital precoding at the large-scale MIMO system is not practical since it requires a huge number of  RF chains which increases the system power consumption and hardware complexity.  Against this background, this paper aims to propose a cost-efficient method, in which a CEP method with passive PSs is employed at the BS.  We aim to minimize the power of multi-user interference (MUI) by jointly optimizing the discrete phases of the BS antennas and the IRS reflective phases.
This optimization problem is first addressed by using the trellis-based method to guarantee the unit modulus constraint of phase shifters at both the BS and the IRS. Then, semi-definite relaxation (SDR)-based method was provided to serve as the performance benchmark method for the proposed trellis-based method. Specifically, the contributions of this paper are summarized as follows:
\begin{itemize}
    \item In this paper,  an IRS-aided multi-cell large-scale MIMO system using the passive constant envelope precoding is proposed, where  discrete PSs are utilized in both the IRS and the BS.
    \item  \textcolor{\revisioncolor}{For the sake of reducing the sum power of MUI terms, an effective approach is devised}  by jointly optimizing the discrete phase shifts at both the BS and the IRS. At first, a trellis-based algorithm is proposed for a single-cell scenario, and then it is extended to a multi-cell scenario, where the precoding operation in each cell is performed individually and simultaneously.
    \item Employing stochastic optimization, a low-overhead trellis-based solution is presented where the precoding vectors of all cells are calculated simultaneously, such that they minimize the inter-cell interference as well as the intra-cell MUI, to further enhance the \textcolor{\revisioncolor}{overall system throughput}.
    \item An SDR-based solution to the optimization problem is also presented, which serves as the performance benchmark for  the low-complexity trellis-based solutions.
    \item We provide overall analysis for computational complexity, overhead and
    interference robustness of the proposed schemes.
\end{itemize}

\textcolor{\revisioncolor}{The remainder of the paper is organized as follows:} the system model and the problem formulation are presented in Section \ref{System Model}. Section \ref{trellis single cell} and \ref{trellis multi cell} are dedicated to the trellis-based algorithm for joint  BS precoding and IRS beamforming for single-cell scenario and multi-cell scenario, respectively. In Section \ref{SDR-based solution for multi cell}, the SDR-based solution is provided, and  interference and complexity analysis for the presented schemes are given in Sections \ref{interference analisys} and \ref{Complexity Analysis}, respectively. Numerical results are given in Section \ref{Numerical Evaluations}, and finally Section \ref{Conclusion} concludes this paper.

Notations : lower case italic letters (e.g. $i$) represent variables, lower case boldface  letters (e.g. $\mb{x}$) represent vectors, upper case italic letters (e.g. $M$) are constants and upper case boldface letters (e.g. $\mb{H}$) denote matrices. The operators $|.|$ $(.)^H$, $(.)^*$ and $(.)^T$ respectively stand for the  first order norm, Hermitian, conjugate and transpose of a vector or matrix. Given a matrix with complex entries and the dimension of $N\times N$, e.g.  $\mb{M}\in\mathbb{C}^{N\times N}$, $\rank{(\mb{M})}$, $\tr{(\mb{M})}$,  and $\mb{M}_{ij}$ respectively stand for the rank of this matrix, its trace,  and the $(i,j)$th entry of $\mb{M}$. When $\mb{M}$ is a positive semi-definite matrix, the notation $\mb{M}\succeq 0$ is used. Also, $\diag{(\mb{M})}$, is an $N\times 1$ vector with all the entries from the diagonal elements of $\mb{M}$, while  $\diag{(\mb{x})}$, $\mb{x}$ being an $N\times 1$ vector, is an $N\times N$ diagonal matrix with $\mb{x}$ being its diagonal elements. The symbols $\mb{0}$ and $\mb{1}$ denote all zero and all one matrices with appropriate sizes. A random vector with complex normal distribution is denoted by $\mb{r}\sim\mathcal{CN}(\mb{v},\mb{\Sigma})$ where $\mb{v}$ and $\mb{\Sigma}$ are its mean and covariance matrix, and $\mb{r}_j$ is the $j$-th element of this vector. Finally, $\mathbb{E}\{.\}$ represents the expectation operator.

\section{System Model and Problem Formulation}
\label{System Model}
In this paper, we consider the downlink transmission of  a  multi-user large-scale MIMO system aided by an IRS. For convenience, we first consider the application of the IRS in a single-cell scenario, and then extend to a multi-cell scenario. It is assumed that CEP is utilized as the precoding scheme at the BSs, where each BS has $N_T$ PSs and one PA with transmit power $P_T$. In this case, the signal transmitted by each antenna has the same amplitude, but different phases. In addition,  the IRS is composed of $M$ PSs, which assists the BSs to transmit their signals to $K$ single-antenna users. 
\begin{figure}[t]
\begin{center}
   \includegraphics[scale=.58]{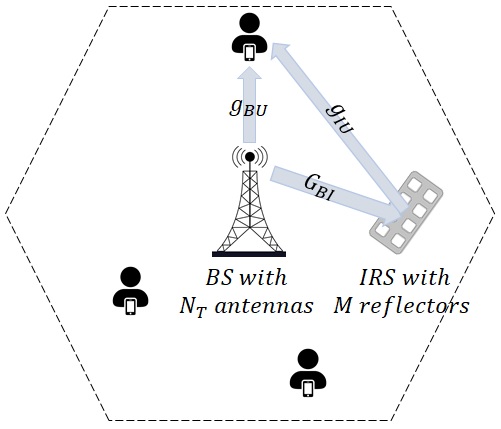}
       \caption{Single-cell system model with $K=3$ single-antenna users. The available channels to a specific user are also demonstrated.}
    \label{fig:single-cell system model}
\end{center}
\end{figure}
For a single-cell model with an $N_T$-antenna BS, an $M$-antenna IRS and $K$ single-antenna users, as shown in Fig. \ref{fig:single-cell system model}, the received signal at the $i$-th user can be written as
\begin{equation}\label{Reciv Sig}
y^{[i]}=\mb{g}_{BU}^{[i]}\mb{x}+\mb{g}_{IU}^{[i]}\Psi_{I}\mb{G}_{BI}\mb{x}+w^{[i]},
\end{equation}
in which $\mb{g}_{BU}^{[i]}$ is the $1\times N_T$ channel vector between the BS and the $i$-th user terminal, $\mb{x}$ stands for \textcolor{\revisioncolor}{the vector of} the CE transmitted signal\textcolor{\revisioncolor}{s} from the BS, $\mb{g}_{IU}^{[i]}$ denotes the $1\times M$ channel vector between the IRS and the $i$-th user terminal, $\Psi_{I}=\diag\left[\psi_1,\dots \psi_M\right]$ denotes the $M\times M$ diagonal matrix of the phase shifters of the IRS,  $\mb{G}_{BI}$ represents the $M\times N_T$ channel matrix between the BS and the IRS, and  $w^{[i]}\sim\mathcal{CN}(0,\sigma_{w}^{})$ is the additive noise at the $i$-th user.
The channel coefficients are modeled as in \cite{Marzetta2010NoncooperativeAntennas} in which each channel comprises the small-scale fading and the large-scale fading coefficients, e.g.
\begin{equation}
    \mb{g}_{BU}^{[i]}=\mb{h}^{[i]}\beta^{[i]\frac{1}{2}},
\end{equation}
where $\mb{h}^{[i]}\sim\mathcal{CN}(\mb{0},\mb{1}_{N_T})$ denotes the small-scale fading coefficients and \textcolor{\revcolor}{$\beta_{}^{[i]}$} is the large-scale fading coefficient. Note that all the other channel coefficients, i.e. $\mb{G}_{BI}$ and $\mb{g}_{IU}^{[i]}$, are modeled similarly.
Assuming that the desired symbol for the $i$-th user is represented by $s^{[i]}$, the MUI at this terminal is given by
\begin{equation}  \label{MUI term for single cell}
    e^{[i]}=\mb{g}_{BU}^{[i]}\mb{x}+\mb{g}_{IU}^{[i]}\Psi_{I}\mb{G}_{BI}\mb{x}-s^{[i]}.
\end{equation}
Hence, the received signal in (\ref{Reciv Sig}) is rewritten as a superposition of the desired symbol, the MUI and the receiver noise as follows
\begin{equation}\label{superposition of ds mui and noise}
    y^{[i]}=s^{[i]}+e^{[i]}+w^{[i]}.
\end{equation}
In order to evaluate the impact of the interference in (\ref{MUI term for single cell}),  the total power of MUI for all users is calculated as follows
\begin{equation}\label{p_e}
    P_{e}^{si}=\sum_{i=1}^{K}\left|e^{[i]}\right|^{2}=\sum_{i=1}^{K}\left|\mb{g}_{BU}^{[i]}\mb{x}+\mb{g}_{IU}^{[i]}\Psi_{I}\mb{G}_{BI}\mb{x}-s^{[i]}\right|^{2}.
\end{equation}
The main idea \textcolor{\revisioncolor}{behind designing  an IRS-aided precoding} is to optimize the phases of the BS antennas and the IRS reflectors such that the power of MUI is minimized. 

In this paper, we aim to jointly design the PSs at both the BS and IRS with the aim of minimizing the power of MUI. Then  for the single-cell scenario, the optimization problem can be reformulated
\begin{equation}\label{main opt}
\begin{array}{cl}
\min\limits_{\mb{x}, \Psi_{I}}& P_{e}^{si}\\ \\
\st
&\mb{x}_{}^{}=\frac{P_{T}}{N_{T}}\left[e^{\jj \theta_{1}},\, e^{\jj \theta_{2}},\dots,\ e^{\jj \theta_{N_{T}}}\right]^T,\\
&\diag\left(\Psi_{I}^{}\right)=\left[e^{\jj \phi_{1}},\, e^{\jj \phi_{2}},\dots,\ e^{\jj \phi_{M}}\right]^T.
\end{array}
\end{equation}
Expanding the objective function in (\ref{p_e})  leads to:
\begin{align} \label{ext p_e}
    P_{e}^{si}=\sum_{i=1}^{K}\Bigg[&\mb{x}^{H}\mb{g}_{BU}^{[i] \ H}\mb{g}_{BU}^{[i]}\mb{x}\nonumber\\&+\mb{x}^{H}\mb{G}_{BI}^{H}\Psi_{I}^{H}\mb{g}_{IU}^{[i] \ H}\mb{g}_{IU}^{[i]}\Psi_{I}\mb{G}_{BI}\mb{x}\nonumber\\&+2\re\left\{\mb{x}^{H}\mb{G}_{BI}^{H}\Psi_{I}^{H}\mb{g}_{IU}^{[i] \ H}\mb{g}_{BU}^{[i]}\mb{x}\right\}\nonumber\\
    &-2\re\left\{\mb{g}_{BU}^{[i]}\mb{x}s^{[i]*}\right\}\nonumber\\&-2\re\left\{\mb{g}_{IU}^{[i]}\Psi_{I}\mb{G}_{BI}\mb{x}s^{[i]*}\right\}+\left|s^{[i]}\right|^2\Bigg].
\end{align}
Note that it is challenging to solve Problem (\ref{ext p_e}) as two optimization variables ($\mb{x}$ and $\Psi_{I}$) are coupled.
Thus, to make it tractable, it is  divided  into two separate sub-problems,
one optimizes $\mb{x}$ while fixing $\Psi_{I}$, and vice versa. In the following, the main problem is reformulated into sub-problems.
\subsection{Optimize $\mb{x}$ with fixed $\Psi_{I}$} \label{subsection A}
In the first stage, by denoting the fixed value of the phase shifts of the IRS as $\Psi_{I}=\Psi_{I}^{\text{Initial}}$   and ignoring the terms independent of $\mb{x}$, the optimization problem in (\ref{main opt}) can be rewritten as

\begin{equation}\label{x opt}
\begin{array}{cl}
\min\limits_{\mb{x}}&
P_x \\ \\
\st
&\mb{x}_{}^{}=\frac{P_{T}}{N_{T}}\left[e^{\jj \theta_{1}},\, e^{\jj \theta_{2}},\dots,\ e^{\jj \theta_{N_{T}}}\right]^T,
\end{array}
\end{equation}
where
\begin{align}\label{P_x}
P_x=&\displaystyle\sum_{j=1}^{N_T}\re\Bigg\{\displaystyle\sum_{i=1}^{j-1}\Big[x_{j}^{*}x_{i}\mb{H}_{BU_{j,i}}+x_{j}^{*}x_{i}\mb{T}_{j,i}\nonumber\\&+x_{j}^{*}x_{i}\left(\mb{Z}_{j,i}+\mb{Z}_{i,j}^{*}\right)\Big]-\mb{s}^{H}\mb{G}_{BU_{j}}x_{j}-\mb{s}^{H}\mb{F}_{j}\mb{x}_{j}\Bigg\},
\end{align}
with
\begin{equation}
    \mb{H}_{BU}\triangleq\mb{G}_{BU}^{H}\mb{G}_{BU},
\end{equation}
\begin{equation}
    \mb{F}\triangleq \mb{G}_{IU}^{}\Psi_{I}^{\text{Initial}}\mb{G}_{BI},
\end{equation}
\begin{equation}
    \mb{T}\triangleq\mb{F}^{H}\mb{F},
\end{equation}
\begin{equation}
    \mb{Z}\triangleq\mb{G}_{BI}^{H}\Psi_{I}^{\text{Initial} \ H}\mb{G}_{IU}^{H}\mb{G}_{BU}.
\end{equation}

Since digital PSs are more advantageous than analog ones in terms of power efficiency \textcolor{\revisioncolor}{and implementation cost}, it is more appealing to deploy  discrete-phase PSs, especially in the cases with a large number of PSs. The power consumption of a PS depends on its type and resolution. In \cite{ROI2016HybridSwitches}, the consumed power for 3-, 4-, 5- and 6-bit phase shift is  15, 45, 60, and 78mW, respectively. Also, in \cite{HuangEnergy2018} \textcolor{\inquirycolor}{it is assumed that} a 1-, 2- and infinite-bit resolution PS consumes 5, 15 and 45 dBm of power. As a result, we aim to solve  discrete optimization problems. In this case, the optimization problem in (\ref{x opt}) is replaced by
\begin{equation}\label{equi x opt}
\begin{array}{cl}
\min\limits_{\mb{x}}& P_x  \\ \\
\st
&x_{j}\in \mathbb{X}_{N_{BS}},\quad j=1, \dots, N_{T},
\end{array}
\end{equation}
where $\mathbb{X}_{N_{BS}}=\left\{\frac{P_{T}}{N_{T}}e^{\frac{\jj2\pi n_{j}}{N_{BS}}}, \quad n_{j}=1,\dots N_{BS}\right\}$ and $N_{BS}$ is the number of possible phases that can be selected by each BS antenna.

\subsection{Optimize $\Psi_{I}$ with fixed $\mb{x}$}
In the second stage, by denoting the fixed value of $\mb{x}$ at  \textcolor{\revisioncolor}{$\mb{x}=\mb{x}^{opt}$}, and  ignoring the terms independent of $\Psi_{I}$, the optimization problem in (\ref{main opt}) can be rewritten as

\begin{equation}\label{psi opt}
\begin{array}{cl}
\min\limits_{\Psi_{I}}& P_{\Psi_I} \\ \\
\st
&\diag\left(\Psi_{I}^{}\right)=\left[e^{\jj \phi_{1}},\,\dots,\ e^{\jj \phi_{M}}\right]^T,
\end{array}
\end{equation}
\\
where
\begin{align}\label{P_sai}
    P_{\Psi_I}=\displaystyle\sum_{j=1}^{M}\re\Bigg\{&\displaystyle\sum_{i=1}^{j-1}\left[v_{j}^{*}\psi_{j}^{*}v_{i}\psi_{i}\mb{Q}_{IU_{j,i}}\right]\nonumber\\&-\mb{s}^{H}\mb{G}_{IU_{j}}v_{j}\psi_{j}+v_{j}^{*}\psi_{j}^{*}u_{j}\Bigg\} ,
\end{align}
with
\begin{equation}
    \mb{Q}_{IU}=\mb{G}_{IU}^{H}\mb{G}_{IU},
\end{equation}
\begin{equation}
    \mb{v}=\left[v_1,\dots,v_M\right]^T=\mb{G}_{BI}\textcolor{\revcolor}{\mb{x}^{opt}},
\end{equation}
\begin{equation}
    \mb{u}=\left[u_1,\dots,u_M\right]^T=\mb{G}_{IU}^{H}\mb{G}_{BU}\textcolor{\revcolor}{\mb{x}^{opt}}.
\end{equation}

\textcolor{\revisioncolor}{Similar to Section \ref{subsection A}}, the optimization problem in (\ref{psi opt}) can be rewritten as
\begin{equation}\label{equi psi opt}
\begin{array}{cl}
\min\limits_{\Psi_{I}}& P_{\Psi_{I}}  \\ \\
\st
&\psi_{j}\in \mathbb{D}_{N_{IRS}},\quad j=1, \dots, M,
\end{array}
\end{equation}
\\
in which $\mathbb{D}_{N_{IRS}}=\left\{e^{\frac{\jj2\pi m_{j}}{N_{IRS}}}, \quad m_{j}=1,\dots N_{IRS}\right\}$ and $N_{IRS}$ denotes the number of possible phases that can be selected by each phase shifter of the IRS.
Note that the optimization problems in (\ref{equi x opt}) and (\ref{equi psi opt}) are challenging to solve, as they are non-convex NP-hard problems with discrete optimization variables.
In the following, we aim to find \textcolor{\revcolor}{tractable} low-complexity solutions for these problems.

\section{Trellis-Based Solution for the Single-Cell Scenario} \label{trellis single cell}
This section aims to devise a low complexity scheme to solve the optimization problems in (\ref{equi x opt}) and (\ref{equi psi opt}).
It is noted that the objective function in (\ref{equi x opt}) is the sum of $N_{T}$ real terms, where the $j$-th term is a function of the first $j$ variables. Also, the objective functions in (\ref{equi psi opt}) is the sum of $M$ real terms as well. Hence, these problems can be treated by utilizing a sequential structure.
\begin{figure}[t]
\begin{center}
   \includegraphics[width=\columnwidth]{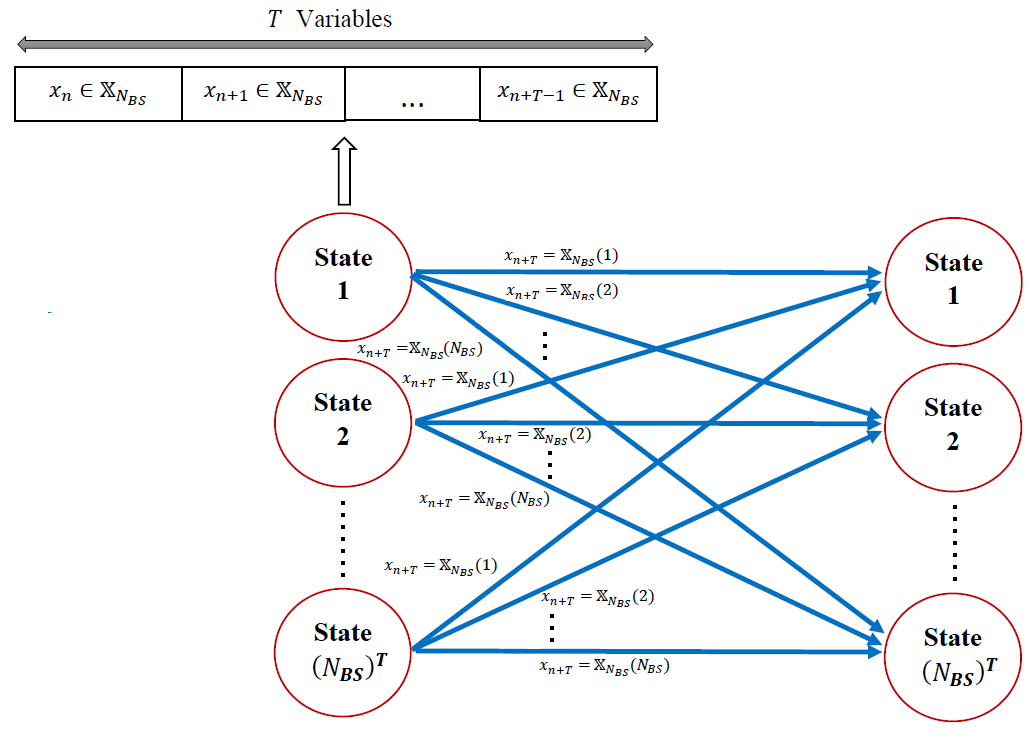}
       \caption{Trellis-based design for determination of $\mb{x}$.}
    \label{fig:my_label x}
\end{center}
\end{figure}

For the sake of solving  Problem (\ref{equi x opt}), let $T$ be the number of variables ($x_i$) as the memory, each having $N_{BS}$ possible choices. Therefore, as depicted in Fig. \ref{fig:my_label x}, we construct a trellis with $(N_{BS})^{T}$ states, each having $N_{BS}$ outgoing branches with labels selected from $\mathbb{X}_{N_{BS}}$.

We describe the details of the proposed algorithm as follows. At  first, the initial $T$ variables are taken as initial memory values and their $(N_{BS})^T$ possible permutations form the trellis states. Initial benchmarks are calculated by inserting the values of each state in the first $T$ terms of the objective function of (\ref{equi x opt}). At the $n$-th stage, the branch labels imply the $(n+T)$-th variable ($x_{n+T}$), and the branch benchmark is the ($n+T$)-th term of the objective function of (\ref{equi x opt}). At each stage, the cumulative benchmark of branches are calculated by adding the branch benchmarks to the cumulative benchmark of their originating paths. After that, among the branches entering the same state, the branch with the least cumulative benchmark is kept and the others are removed. The algorithm is terminated after $N_{T}-T$ stages and the path with the minimum cumulative benchmark is selected as the final solution. The labels on the selected path represent the latter $N_{T}-T$ variables, and the initial state associated with the selected path stands for the first $T$ variables.
\begin{figure}[t]  \label{Trellis-based design for psi}
\begin{center}
   \includegraphics[width=\columnwidth]{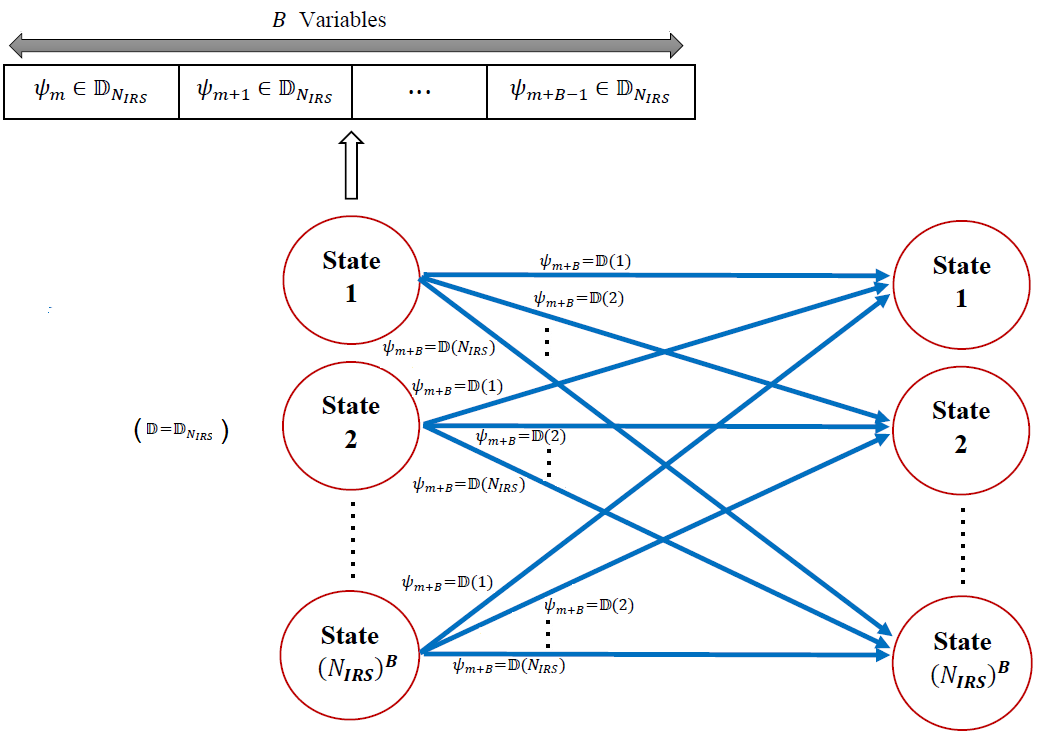}
       \caption{Trellis-based design for determination of $\Psi_{I}$}
    \label{fig:my_label}
\end{center}
\end{figure}

The same procedure could be adopted for solving the optimization problem in (\ref{equi psi opt}) and (\ref{multi cell equi psi opt}) by replacing $N_{BS}$ and $\mathbb{X}_{N_{BS}}$ with $N_{IRS}$ and $\mathbb{D}_{N_{IRS}}$ as illustrated in Fig. \ref{fig:my_label}. To this end,  at  first, the initial $B$ variables are taken as initial memory values and their $(N_{IRS})^{B}$ possible combinations form the trellis states. Initial benchmarks are calculated by inserting the values of each state in the first $B$ terms of the objective function of (\ref{equi psi opt}). At the $m$-th stage, the branch labels imply the $(m+B)$-th variable ($\psi_{m+B}$), and the branch benchmark is the ($m+B$)-th term of the objective function of (\ref{equi psi opt}). At each stage, the cumulative benchmark of branches are calculated by adding the branch benchmarks to the cumulative benchmark of their originating paths. After that, among the branches entering the same state, the branch with the least cumulative benchmark is kept and the others are removed. The algorithm is terminated after $M-B$ stages and the path with the minimum cumulative benchmark is selected as the final solution. The labels on the selected path represent the latter $M-B$ variables, and the initial state corresponding to the selected path stands for the first $B$ variables.

In order to solve the main optimization problem in (\ref{main opt}), at first, assuming a fixed $\Psi_{I}$ (by selecting an initial value for $\Psi_{I}$)  and a variant $\mb{x}$, the optimization problem in (\ref{equi x opt}) is solved according to the proposed algorithm. In the following, based on the optimum value of $\mb{x}$ obtained in the previous stage, the optimization problem in (\ref{equi psi opt}) is solved by assuming a fixed $\mb{x}$ and a variant $\Psi_{I}$. The overall algorithm is presented in Algorithm 1.

\begin{algorithm} \label{Alg1}
\caption{Trellis-based joint BS precoding and IRS beamforming design for a single-cell scenario}
\begin{algorithmic}
\STATE $\mb{Input}:$ $\mathbb{X}_{N_{BS}},\mathbb{D}_{N_{IRS}},\Psi_I^{Initial}$
\STATE $\mb{Output}:$ $\Psi_I, \mb{x}$
\STATE $\mb{Initialize}$ all possible permutations of $x_1,\dots,x_T$
\FOR{$i=T+1,\dots,N_T$}
\FOR{$j=1,\dots,N_{BS}$}
\STATE $x_i=\mathbb{X}_{N_{BS}}(j)$;
\STATE calculate (\ref{P_x}) as the benchmark;
\ENDFOR
\STATE eliminate all paths except the one with the minimum benchmark value;
\ENDFOR
\STATE Choose $x_i,\dots,x_{N_T}$, such that they lead to the minimum cumulative benchmark value.
\STATE $\mb{Initialize}$ all possible permutations of $\psi_{1},\dots,\psi_{B}$
\FOR{$i=B+1,\dots,M$}
\FOR{$j=1,\dots,N_{IRS}$}
\STATE $\psi_{j}=\mathbb{D}_{N_{IRS}}(j)$;
\STATE calculate (\ref{P_sai}) as the benchmark;
\ENDFOR
\STATE eliminate all paths except the one with the minimum benchmark value;
\ENDFOR
\STATE Choose $\psi_{1},\dots,\psi_{M}$, such that they lead to the minimum cumulative benchmark value.
\end{algorithmic}
\end{algorithm}

\section{Trellis-Based Solution for the Multi-Cell Scenario} \label{trellis multi cell}
\begin{figure}[t]
\begin{center}
   \includegraphics[scale=.56]{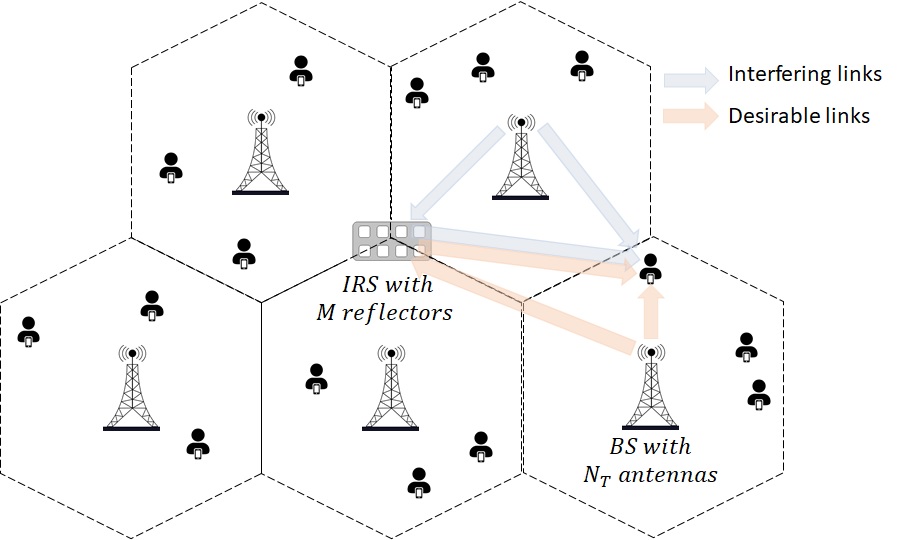}
       \caption{Multi-cell system model with $L=5$ cells each containing a BS and $K=3$ single-antenna users. The interfering and the desirable links for a specific user are also illustrated.}
    \label{fig:multi-cell system model}
\end{center}
\end{figure}
Here, we extend our work to the multi-cell scenario shown  in Fig. \ref{fig:multi-cell system model} where there are $L$ cells, each of which is equipped  with an IRS with $M$ PSs. Then, the received signal at the $i$-th user of the $j$-th cell (referred to as the target user) is given by
\begin{equation} \label{multi-cell received signal}
    y^{[i,j]}=\sum_{l=1}^{L}\mb{g}_{BU_l}^{[i,j]}\mb{x}_l+\sum_{l=1}^{L}\mb{g}_{IU}^{[i,j]}\Psi_{I}\mb{G}_{BI_l}\mb{x}_l+w^{[i,j]},
\end{equation}
where $\mb{g}_{BU_l}^{[i,j]}$ is the $1\times N_T$ channel vector between the $l$-th BS and the $i$-th user in the $j$-th cell, $\mb{x}_l$ denotes the precoding vector of the $l$-th BS, $\mb{g}_{IU}^{[i,j]}$ stands for the $1\times M$ channel vector between the IRS and the target user, $\mb{G}_{BI_l}\in \mathbb{C}^{M\times N_T}$ is the channel  matrix between the $l$-th BS and the IRS and $w^{[i,j]}\sim\mathcal{CN}(0,\sigma_{w}^{})$ represents the additive noise at the target user.
The MUI term for the target user in this case is written as
\begin{equation} \label {multi-cell MUI}
    e^{[i,j]}=\sum_{l=1}^{L}\mb{g}_{BU_l}^{[i,j]}\mb{x}_l+\sum_{l=1}^{L}\mb{g}_{IU}^{[i,j]}\Psi_{I}\mb{G}_{BI_l}\mb{x}_l-s^{[i,j]},
\end{equation}
where $s^{[i,j]}$ is the desired symbol for the target user. Similarly, the power of MUI for the multi-cell scenario is given by
\begin{align} \label {multi-cell Power of MUI}
    P_e^{mu}&=\sum_{j=1}^{L}\sum_{i=1}^{K}\left|e^{[i,j]}\right|^{2}\nonumber\\&=\sum_{j=1}^{L}\sum_{i=1}^{K}\left|\sum_{l=1}^{L}\mb{g}_{BU_l}^{[i,j]}\mb{x}_l+\sum_{l=1}^{L}\mb{g}_{IU}^{[i,j]}\Psi_{I}\mb{G}_{BI_l}\mb{x}_l-s^{[i,j]}\right|^{2}.
\end{align}
Hence, the corresponding optimization problem is formulated as
\begin{equation}\label{multi-cell main opt}
\begin{array}{cl}
\min\limits_{\Psi_{I},\mb{x}_l }& P_{e}^{mu}\\ \\
\st
&\mb{x}_{l}^{}=\frac{P_{T}}{N_{T}}\left[e^{\jj \theta_{1}},\, e^{\jj \theta_{2}},\dots,\ e^{\jj \theta_{N_{T}}}\right]^T,\\ \\ &l=1,\dots,L\\  \\
&\Psi_{I}^{}=\diag\left(\left[e^{\jj \phi_{1}},\, e^{\jj \phi_{2}},\dots,\ e^{\jj \phi_{M}}\right]^T\right),
\end{array}
\end{equation}
Expanding the objective function in  (\ref{multi-cell Power of MUI}) leads to:
\begin{align} \label{multi-cell ext p_e}
    P_{e}^{mu}=&\sum_{j=1}^L\sum_{i=1}^{K}\Bigg[\sum_{l=1}^L\sum_{l^{\prime}=1}^L\mb{x}_l^{H}\mb{g}_{BU_l}^{[i,i] \ H}\mb{g}_{BU_{l^{\prime}}}^{[i,j]}\mb{x}_{l^{\prime}}\nonumber\\&+\sum_{l=1}^L\sum_{l^{\prime}=1}^L\mb{x}_l^{H}\mb{G}_{BI_l}^{H}\Psi_{I}^{H}\mb{g}_{IU}^{[i,j] \ H}\mb{g}_{IU}^{[i,j]}\Psi_{I}\mb{G}_{BI_{l^{\prime}}}\mb{x}_{l^{\prime}}\nonumber\\&+2\re\left\{\sum_{l=1}^L\sum_{l^{\prime}=1}^L\mb{x}_l^{H}\mb{G}_{BI_l}^{H}\Psi_{I}^{H}\mb{g}_{IU}^{[i,j] \ H}\mb{g}_{BU_{l^{\prime}}}^{[i,j]}\mb{x}_{l^{\prime}}\right\}\nonumber\\
    &-2\re\left\{\sum_{l=1}^{L}\mb{g}_{BU_l}^{[i,j]}\mb{x}_ls^{[i,j]*}\right\}\nonumber\\&-2\re\left\{\sum_{l=1}^{L}\mb{g}_{IU}^{[i,j]}\Psi_{I}\mb{G}_{BI_l}\mb{x}_ls^{[i,j]*}\right\}+\left|s^{[i,j]}\right|^2\Bigg].
\end{align}
\subsection{Main trellis-based solution}
For the multi-cell scenario with $L$ cells sharing one IRS,  we first assume  \textcolor{\revisioncolor}{ that} each cell performs its precoding  separately as in the single-cell scenario.
In other words, each BS solves (\ref{x opt}) independently to determine its precoding vector.
However, the phase shift optimization of the IRS is completely different. The IRS requires the precoding vectors of all cells and all the inter-cell channel coefficients to determine its  phase shifts.
The phase shifts of the IRS are designed  to minimize the power of MUI in (\ref{multi-cell ext p_e}). To this end, by considering $\mb{x}_l$, $l=1,\dots,L$, to be constant and $\Psi_I$ to be variables, and by removing the terms independent of $\Psi_I$, (\ref{multi-cell ext p_e}) can be rewritten as

\begin{align}\label{multi-cell P_Sai_2}
        P_{\Psi_I}^{mu}=\displaystyle\sum_{s=1}^{M}\re\Bigg\{&\displaystyle\sum_{s^{\prime}=1}^{s-1}\left[a_s^{*}\psi_s^{*}a_{s^{\prime}}\psi_{s^{\prime}}\omega_{ss^{\prime}} \right]\nonumber\\&-\psi_sa_s\tr\left\{\mb{H}_s\mb{S}^{H}\right\}\nonumber\\&+a_s^*\psi_s^*\tr\left\{\mb{H}_s^{H}\mb{B}\right\}\Bigg\} ,
\end{align}
in which $\mb{H}_{s}$ and $\mb{B}$ denote  $L\times K$ matrices comprising the following entries, respectively
\begin{align}
      &h_{s_{j,i}}\triangleq\mb{g}_{IU_{s}}^{[i,j]},\\
      &b_{j,i}\triangleq\sum_{l=1}^{L}\mb{g}_{BU_{l^{}}}^{[i,j]}\mb{x}_l, \ i=1,\dots,K, \ j=1,\dots L.
\end{align}
In (\ref{multi-cell P_Sai_2}),  $a_i$ is the $i$-th element of the $M\times 1$ vector $\mb{a}=\sum_{l=1}^L\mb{G}_{BI_l}\mb{x}_l$, $\omega_{ij}$ is the element in the $i$-th row and the $j$-th column of the matrix $\mb{W}=\sum_{j=1}^L\sum_{i=1}^K\mb{g}_{IU}^{[i,j] \ H}\mb{g}_{IU}^{[i,j]}$,
and $\mb{S}$ is the $L\times K$ matrix of all desired symbols.

Considering a set of discrete phases for the IRS reflectors, the optimization problem for the multi-cell scenario can be formulated as
\begin{equation}\label{multi cell equi psi opt}
\begin{array}{cl}
\min\limits_{\Psi_{I}}& P_{\Psi_{I}}^{mu}  \\ \\
\st
&\psi_{j}\in \mathbb{D}_{N_{IRS}},\quad j=1, \dots, M.
\end{array}
\end{equation}
\textcolor{\revisioncolor}{As mentioned before, it is evident that the optimization problem in (\ref{multi cell equi psi opt}) is available once all BSs have computed their precoding operations.}

Note that all channel coefficients are required in this scenario, which incurs high CSI exchange or estimation overhead.
In the following section, we attempt to reduce the imposed overhead on the system, while using all of the available CSI in the BS precoding.
\subsection{Low-overhead trellis-based solution}
Here, we aim to reduce the amount of imposed overhead on the system. To this end, we adopt the following assumptions:
\begin{itemize}
    \item The intra-cell channel coefficients between each BS and its users are known, i.e. $\mb{g}_{BU_j}^{[i,j]}$, $i=1,...,K$.
    \item The channel coefficients between each BS and the IRS are known, i.e. $\mb{G}_{BI_l}$, $l=1,...,L$.
    \item \textcolor{\inquirycolor}{The channel coefficients between the IRS and the users are known, i.e. $\mb{g}_{IU}^{[i,j]}$, for all $i,j$.}
    \item The channel coefficients between a BS and the users in other cells, i.e. $\mb{g}_{BU_l}^{[i,j]}$, $l\neq j$ , are not available, but their first and second order statistical information is available.
\end{itemize}
Hence, the overhead is reduced by $L\times (L-1)\times K\times N_T$, which is a considerable amount, especially in large-scale MIMO systems.

Now by adopting the stochastic optimization method \cite{Shapiro2009}, we aim to derive the power of the MUI term in (\ref{multi-cell P_Sai_2}), based on which we formulate the following optimization problem
\begin{equation}\label{main stoch. opt.}
\begin{array}{cl}
\min\limits_{\mb{x}}&
f_{0}(\mb{x},g_1,\dots,g_N) \\ \\
\st
& f_{i}(\mb{x},g_1,\dots,g_N)= 0, \ i=1,\dots, M.
\end{array}
\end{equation}
As $g_1,\dots, g_N$ are random variables, the optimization problem
in (\ref{main stoch. opt.}) can be approximated as
\begin{equation}\label{eqiv. stoch. opt.}
\begin{array}{cl}
\min\limits_{\mb{x}}&
\mathbb{E}_{g_1,\dots\,g_N}\left[f_{0}(\mb{x},g_1,\dots,g_N)\right] \\ \\
\st
& \mathbb{E}_{g_1,\dots\,g_N}\left[f_{i}(\mb{x},g_1,\dots,g_N)\right]= 0, \ i=1,\dots, M.
\end{array}
\end{equation}
Now, employing the stochastic optimization and  assuming
\begin{equation}
   \mb{g}_{BU_l}^{[i,j]}\approx\mathbb{E}\left\{\mb{g}_{BU}^{[i,j]}\right\}=\bs{0},
\end{equation}
\begin{equation}
    \mb{g}_{BU_l}^{[i,j] \ H}\mb{g}_{BU_l}^{[i,j]}\approx\mathbb{E}\left\{\mb{g}_{BU_l}^{[i,j] \ H}\mb{g}_{BU_l}^{[i,j]}\right\}\approx\beta_{BU_l}^{[i,j]}\bs{I}_{N_T},
\end{equation}
in which $\beta_{BU_l}^{[i,j]}$ is the large-scale fading component of $\mb{g}_{BU_l}^{[i,j]}$, the power of MUI for the IRS is determined as
\begin{align}\label{multi-cell P_Sai_3}
        P_{\Psi_I}^{mu}\approx\displaystyle\sum_{s=1}^{M}\re\Bigg\{&\displaystyle\sum_{s^{\prime}=1}^{s-1}\left[a_s^{*}\psi_s^{*}a_{s^{\prime}}\psi_{s^{\prime}}\omega_{ss^{\prime}} \right]\nonumber\\&-\psi_sa_s\tr\left\{\mb{H}_s\mb{S}^{H}\right\}\nonumber\\&+a_s^*\psi_s^*\tr\left\{\mb{H}_s^{H}\mb{B}^{\prime}\right\}\Bigg\} ,
\end{align}
where $\mb{B}^{\prime}$ denotes an $L\times K$ matrix with the entries $b^{\prime}_{j,i}=\mb{g}_{BU_j}^{[i,j]}\mb{x}_j$, $i=1,\dots K$ and $j=1\dots L$. By solving the optimization problem in (\ref{multi cell equi psi opt}), the phase shifts of the IRS are determined.

Now, to achieve better performance, a new precoding algorithm is presented for the BSs, where each BS considers not only the MUI term but also the inter-cell interference terms. To this end, motivated by \cite{Shahabi2019IET},  multiplying the MUI term in (\ref{multi-cell MUI}) by $\frac{1}{N_T}\mb{x}_j^H\mb{x}_j=1$ leads to the following expression for \textcolor{\revisioncolor}{the MUI term of the $i$-th user in the $j$-th cell (referred to as the target user)}
\begin{align} \label {Dist. MUI}
    e^{[i,j]}=&\mb{g}_{BU_j}^{[i,j]}\mb{x}_j+\frac{1}{N_T}\sum_{\substack{l=1\\l\neq j}}^{L}\mb{g}_{BU_l}^{[i,j]}\mb{\Gamma}_{lj}\mb{x}_j+\mb{g}_{IU}^{[i,j]}\Psi_{I}\mb{G}_{BI_j}\mb{x}_j\nonumber\\&+\frac{1}{N_T}\sum_{\substack{l=1\\l\neq j}}^{L}\mb{g}_{IU}^{[i,j]}\Psi_{I}\mb{G}_{BI_l}\mb{\Gamma}_{lj}\mb{x}_j-s^{[i,j]},
\end{align}
where
\begin{equation}
    \mb{\Gamma}_{lj}=\mb{x}_l\mb{x}_j^H=\begin{bmatrix}
    e^{j(\theta_{1,l}-\theta_{1,j})}&\dots&e^{j(\theta_{1,l}-\theta_{N_T,j})}\\ \vdots& \ddots&\vdots\\e^{j(\theta_{N_T,l}-\theta_{1,j})}&\dots&e^{j(\theta_{N_T,l}-\theta_{N_T,j})}
    \end{bmatrix}.
\end{equation}

By adopting the stochastic optimization,  we  derive the power of the MUI term in (\ref{Dist. MUI}). The power of MUI for the target cell is determined as
\begin{align}\label{Dist. Pmui}
    P_{MUI}^j&=\sum_{i=1}^{K}\Bigg(\mb{x}_j^H\mb{g}_{BU_j}^{[i,j]\ H}\mb{g}_{BU_j}^{[i,j]}\mb{x}_j+2\re\left\{\mb{x}_j^H\mb{g}_{BU_j}^{[i,j]\ H}s^{[i,j]}\right\}\nonumber\\&+2\re\left\{\mb{x}_j^H\mb{g}_{BU_j}^{[i,j]\ H}\mb{g}_{IU}^{[i,j]}\Psi_{I}\mb{G}_{BI_j}\mb{x}_j\right\}\nonumber\\&+\frac{1}{N_T^2}\sum_{\substack{l=1\\l\neq j}}^{L}\mb{x}_j^H\mb{1}_{N_T}\mathbb{E}\left\{\mb{g}_{BU_l}^{[i,j]\ H}\mb{g}_{BU_l}^{[i,j]}\right\}\mb{1}_{N_T}\mb{x}_j\nonumber\\&{+}\frac{1}{N_T^2}\sum_{\substack{l=1\\l\neq j}}^{L}\mb{x}_j^H\mb{1}_{N_T}\mb{G}_{BI_l}^H\Psi_{I}^H \mb{g}_{IU}^{[i,j]\ H}\mb{g}_{IU}^{[i,j]}\Psi_{I}\mb{G}_{BI_l}\mb{1}_{N_T}\mb{x}_j\nonumber\\&+\mb{x}_j^H\mb{G}_{BI_j}^H\Psi_{I}^H\mb{g}_{IU}^{[i,j]\ H}\mb{g}_{IU}^{[i,j]}\Psi_{I}\mb{G}_{BI_j}\mb{x}_j\nonumber\\&-2\re\left\{\mb{x}_j^H\mb{G}_{BI_j}^{H}\Psi_{I}^H\mb{g}_{IU}^{[i,j]\ H}s^{[i,j]} \right\}+\left|s^{[i,j]}\right|^2\Bigg).
\end{align}
Note that $\Gamma_{lj}$ is estimated by an all-one $N_T\times N_T$ matrix, represented by $\mb{1}_{N_T}$. Based on our extensive experiments, $\theta_{m^{\prime}l}-\theta_{mj}$ is a random variable with uniform distribution within $[-\pi, \pi)$, and it can be approximated by its zero mean.
By expanding (\ref{Dist. Pmui}) and removing the terms that are independent of  $\mb{x}_j$, the optimization problem that the target BS needs to solve  can be formulated as
\begin{equation}\label{Dist. x opt}
\begin{array}{cl}
\min\limits_{\mb{x}_j}&
P_{MUI_x}^j \\ \\
\st
&x_{j}\in \mathbb{X}_{N_{BS}},\quad j=1, \dots, N_{T},
\end{array}
\end{equation}
where
\begin{align}\label{Dist. P_MUI_X}
    P_{MUI_x}^j=\sum_{s=1}^{N_T}\re&\Bigg\{\sum_{s^{\prime}=1}^{s-1}x_{js}^*x_{js^{\prime}}\Big(\mb{D}_{ss^{\prime}}+\mb{E}_{s^{\prime}s}+\mb{E}_{ss^{\prime}}^*\nonumber\\&+\mb{N}_{ss^{\prime}}+\frac{\mb{J}_{ss^{\prime}}+\mb{R}_{ss^{\prime}}}{N_T^2} \Big)\nonumber\\&+ x_{js}^*\left(\mb{q}_s-\mb{m}_s \right) \Bigg\},
\end{align}
\begin{equation}
    \mb{D}=\sum_{i=1}^{K}\mb{g}_{BU_j}^{[i,j]\ H}\mb{g}_{BU_j}^{[i,j]},
\end{equation}
\begin{equation}
    \mb{E}=\sum_{i=1}^{K}\mb{g}_{BU_j}^{[i,j]\ H}\mb{g}_{IU}^{[i,j]}\Psi_{I}\mb{G}_{BI_j},
\end{equation}
\begin{equation}
    \mb{N}=\sum_{i=1}^{K}\mb{G}_{BI_j}^{H}\Psi_{I}^H\mb{g}_{IU}^{[i,j]\ H}\mb{g}_{IU}^{[i,j]}\Psi_{I}\mb{G}_{BI_j},
\end{equation}
\begin{equation}
    \mb{J}=\sum_{i=1}^{K}\sum_{\substack{l=1\\l\neq j}}^{L}\mb{\Gamma}_{lj}^H\mathbb{E}\left\{\mb{g}_{BU_l}^{[i,j]\ H}\mb{g}_{BU_l}^{[i,j]} \right\}\mb{\Gamma}_{lj},
\end{equation}
\begin{equation}
    \mb{R}=\sum_{i=1}^{K}\sum_{\substack{l=1\\l\neq j}}^{L}\mb{\Gamma}_{lj}^H\mb{G}_{BI_l}^H\Psi_{I}^H\mb{g}_{IU}^{[i,j]\ H}\mb{g}_{IU}^{[i,j]}\Psi_{I}\mb{G}_{BI_l}\mb{\Gamma}_{lj}.
\end{equation}
\begin{equation}
    \mb{q}=\sum_{i=1}^{K}\mb{g}_{BU_j}^{[i,j] \ H}s^{[i,j]}
\end{equation}
\begin{equation}
    \mb{m}=\sum_{i=1}^{K}\mb{G}_{BI_j}^{H}\bs{\Psi}_{I}^{H}\mb{g}_{IU}^{[i,j] \ H}s^{[i,j]}.
\end{equation}

Note that in this optimization problem, not only the imposed overhead is reduced , but also the interference from other cells is minimized, which leads to better performance.

\section{SDR-based joint BS Precoding and IRS Beamforming Design} \label{SDR-based solution for multi cell}
In this section, the SDR method is used for joint BS precoding and IRS beamforming with discrete-phase PSs. \textcolor{\revisioncolor}{The purposes of this study are twofold;} \textcolor{\inquirycolor}{firstly, the SDR algorithm has not been used for joint discrete IRS beamforming and BS CEP}; \textcolor{\revisioncolor}{ secondly, the SDR algorithm can serve as the performance benchmark for the proposed trellis-based algorithms. }  To this end, we utilize the SDR method to first find the BS precoding vectors by solving the optimization problem in (\ref{equi x opt}), and then determine the IRS phase shifts. By reformulating (\ref{multi-cell MUI}) in a vector form and assuming $\mb{s}^{[j]}$ to be the vector of desired symbols in the $j$-th cell, the vector of MUI terms for the users in the $j$-th cell is written as
\begin{align}
   \mb{e}^{[j]}=\bs{\lambda}^{[j]}\mb{x}_j-\mb{s}^{[j]},
\end{align}
\begin{align}
    \bs{\lambda}^{[j]}=&\mb{G}_{BU_j}^{[j]}+\mb{G}_{IU}^{[j]}\Psi_{I}\mb{G}_{BI_j}\nonumber\\&+\sum_{\substack{l=1\\l\neq j}}^{L}\left(\mb{G}_{BU_l}^{[j]}+\mb{G}_{IU}^{[j]}\Psi_{I}\mb{G}_{BI_l} \right)\mb{\Xi}_{lj},
\end{align}
where
\textcolor{\revcolor}{\begin{equation}
    \mb{G}_{BU}^{[j]}=\left[{\mb{g}_{BU}^{[1j]}}^T,\dots,{\mb{g}_{BU}^{[Kj]}}^T\right]^T,
\end{equation}
\begin{equation}
    \mb{G}_{IU}^{[j]}=\left[{\mb{g}_{IU}^{[1j]}}^T,\dots,{\mb{g}_{IU}^{[Kj]}}^T\right]^T,
\end{equation}}
and $\mb{\Xi}_{lj}=\frac{\mb{x}_l\mb{x}_j^H}{N_T}$ can be estimated by $\frac{1}{N_T}\mb{1}_{N_T}$. In this case, the power of MUI for the $j$-th cell can be represented by
\begin{align}
    P_{MUI}^{[j]}=\left|\left|\mb{e}^{[j]}\right|\right|^2=\mb{x}_j^H\mb{\Lambda}^{[j]}\mb{x}_j-\bs{\xi}^{[j]\ H}\mb{x}_j-\mb{x}_j^H\bs{\xi}^{[j]}+E_{s}^{[j]},
\end{align}
where
\begin{align}
    \mb{\Lambda}^{[j]}=\bs{\lambda}^{[j]\ H}\bs{\lambda}^{[j]},\\
    \bs{\xi}^{[j]}=\bs{\lambda}^{[j]\ H}\mb{s}^{[j]},\\E_{s}^{[j]}=\left|\left|\mb{s}^{[j]}\right|\right|^2,
\end{align}
and the main optimization problem for the $j$-th BS to optimize its transmission phases becomes
\begin{equation}\label{main non convex opt}
\begin{array}{cl}
\min\limits_{\mb{x}_j}& \mb{x}_j^H\mb{\Lambda}^{[j]}\mb{x}_j-\bs{\xi}^{[j]\ H}\mb{x}_j-\mb{x}_j^H\bs{\xi}^{[j]}  \\ \\
\st
& \diag{(\mb{x}_j\mb{x}_j^H)}=\frac{P_{T}}{N_{T}}\mb{1}_{N_T\times 1}.
\end{array}
\end{equation}

The optimization problem in (\ref{main non convex opt}) is non-convex; however, by adopting the SDR method it can be approximated as a convex problem.
By defining
\begin{equation}
    \mb{x}_j^{\prime}=[\mb{x}_j^T,\frac{P_{T}}{N_{T}}]^T,
\end{equation}
and
\begin{equation}
    \mb{\Upsilon}^{[j]}= \left[
    \begin{array}{cc}
        \mb{\Lambda}^{[j]} &  -\bs{\xi}^{[j]} \\
        -\bs{\xi}^{[j]\ H}
          & 0
    \end{array}\right],
\end{equation}
the equivalent optimization problem can be written as
\begin{equation}\label{equi non convex opt}
\begin{array}{cl}
\min\limits_{\mb{x}_j^{\prime}}& \tr\left(\mb{\Upsilon}^{[j]}\mb{x}_j^{\prime}\mb{x}_j^{\prime\ H}\right)  \\ \\
\st
& \diag{(\mb{x}_j^{\prime}\mb{x}_j^{\prime \ H})}=\frac{P_{T}}{N_{T}}\mb{1}_{(N_T+1)\times 1}.
\end{array}
\end{equation}
Now by defining a new variable $\mb{X}^{\prime}_j=\mb{x}_j^{\prime}\mb{x}_j^{\prime \ H}$,
the optimization problem can be rewritten as
\begin{equation}\label{equi2 non convex opt}
\begin{array}{cl}
\min\limits_{\mb{X}_j^{\prime}}& \tr\left(\mb{\Upsilon}^{[j]}\mb{X}_j^{\prime}\right)  \\ \\
\st
& \diag{(\mb{X}_j^{\prime})}=\frac{P_{T}}{N_{T}}\mb{1}_{(N_T+1)\times 1} \\ \\
&\rank(\mb{X}_j^{\prime})=1,\\ \\
&\mb{X}_j^{\prime}\succeq 0.
\end{array}
\end{equation}
If we relax the optimization problem in (\ref{equi2 non convex opt}) by removing the rank one constraint, the following  problem is obtained
\begin{equation}\label{Relaxed equi2 convex opt}
\begin{array}{cl}
\min\limits_{\mb{X}_j^{\prime}}& \tr\left(\mb{\Upsilon}^{[j]}\mb{X}_j^{\prime}\right)  \\ \\
\st
& \mb{X}_{jj}^{\prime}=\frac{P_{T}}{N_{T}}, \ j=1,\dots N_T+1 \\
 \\
&\mb{X}_j^{\prime}\succeq 0.
\end{array}
\end{equation}
The resultant optimization problem in (\ref{Relaxed equi2 convex opt}) is linear and can be readily solved  by the CVX toolbox in Matlab. After obtaining the solution, $\mb{X}^{\prime \ opt}_{j}$, if it is a rank-one matrix,  the optimum BS precoding vector can be found  as
\begin{equation}
    \mb{x}^{\prime \ opt}_{j}=\sqrt{\mu}\mb{k},
\end{equation}
where $\mb{k}$ and $\mu$ are the eigenvector and the eigenvalue of $\mb{X}^{\prime \ opt}_{j}$, respectively. Afterwards, the obtained $\mb{X}^{\prime \ opt}_{j}$ is  mapped into the discrete-phase space. However, if the rank of this matrix is  $r>1$, $\mb{X}^{\prime \ opt}_{j}$
 only gives a lower bound on the objective function. Hence, it must be projected into the constraint space.
  To this end, the following method is presented based on the algorithm of Goemans and Williamson  \cite{Goemans2004}.
  \begin{enumerate}
      \item Solve the SDP relaxation problem in (\ref{Relaxed equi2 convex opt})  to obtain the optimal solution $\mb{X}^{\prime \ opt}_{j}$. Since
$\mb{X}^{\prime \ opt}_{j}$ is a positive semi-definite matrix,  the Cholesky decomposition could be utilized as
$\mb{X}^{\prime \ opt}_{j}=\mb{Z}_j\mb{Z}_j^{H}$. 
      \item Generate a random vector $\mb{r}$ with the distribution of $\mb{r}\sim\mathcal{CN}(\mb{0},\mb{I}_{N_T+1})$.
      \item For $s=1, 2,\dots, N_T+1$, let $\hat{x}_{j,s} = f ({\mb{z}}_{j,s}\times\mb{r})$, where ${\mb{z}}_{j,s}$ denotes the $s$-th row of $\mb{Z}_{j}$ and the function $f (.)$ is defined as follows
  \end{enumerate}
  \begin{align}
          &f(x)=\nonumber\\
          &\begin{cases}
          \frac{P_T}{N_T}, \ \quad\quad\quad\quad\quad \text{arg}(x)\in [  \frac{-\pi}{N_{BS}}
          ,\frac{\pi}{N_{BS}})\\ \frac{P_{T}}{N_{T}}e^{\frac{\jj2\pi}{N_{BS}}}, \ \ \quad\quad\ \text{arg}(x)\in [  \frac{\pi}{N_{BS}}
          ,\frac{3\pi}{N_{BS}})\\ \vdots \\ \frac{P_{T}}{N_{T}}e^{\frac{\jj2\pi (N_{BS}-1)}{N_{BS}}}, \ \text{arg}(x)\in [  \frac{(2N_{BS}-3)\pi}{N_{BS}}
          ,\frac{(2N_{BS}-1)\pi}{N_{BS}}).
          \end{cases}
      \end{align}
  Note that $\hat{x}_{j,s}\in \mathbb{X}_{N_{BS}},  \  \text{for} \  s=1, \dots, N_T+1$, satisfies the constraint of the optimization problem in (\ref{equi non convex opt}). In this case,  $\mb{\hat{x}}_j=[\hat{x}_{j,1}, \dots , \hat{x}_{j,N_T}]$ would be the final solution. \textcolor{\revcolor}{It is worth noting that repeating the algorithm for different values of $\mb{r}$ and selecting the corresponding $\hat{\mb{x}}_{j}$ vector with the lowest objective function value in (\ref{equi non convex opt}) can result in  a more accurate solution.}
  It can be shown that this algorithm results in an $\frac{(N_{BS}\sin(\frac{\pi}{N_{BS}}))^2}{2N_{BS}}$-approximation error \cite{So2007}.

  Now, the same method can be applied for the IRS beamforming as well. To this end, assuming that the obtained values of $\mb{x}_j$ are known at the IRS, the vector of MUI terms for the users in the $j$-th cell is represented by
  \begin{equation}\label{Int term IRS}
      \mb{e}^{[j]}=\bs{\Pi}^{[j]}\mb{\Psi}_{I}^{\vect}-\bs{\eta}^{[j]},
  \end{equation}
  in which
  \begin{equation}
      \bs{\eta}^{[j]}=\mb{s}^{[j]}-\sum_{l=1}^{L}\mb{G}_{BU_l}^{[j]}\bs{x}_l,
  \end{equation}
\begin{equation}
    \bs{\Pi}^{[j]}=\mb{G}_{IU}^{[j]} \diag \left(\sum_{l=1}^{L}\mb{G_{BI_l}}\mb{x}_l\right),
 \end{equation}
\begin{equation}
    \mb{\Psi}_{I}^{\vect}=\diag(\mb{\Psi}_I).
\end{equation}
In addition, the vector of MUI terms for all users is written as
\begin{equation}
    \mb{e}=\left[{\mb{e}^{[1]}}^T,\dots,{\mb{e}^{[L]}}^T\right]^T=\bs{\Pi}\mb{\Psi}_{I}^{\vect}-\bs{\eta},
\end{equation}
where
\begin{equation}
    \bs{\Pi}=\left[{\bs{\Pi}^{[1]}}^T,\dots,{\bs{\Pi}^{[L]}}^T\right]^T,
\end{equation}
\begin{equation}
    \bs{\eta}=\left[{\bs{\eta}^{[1]}}^T,\dots,{\bs{\eta}^{[L]}}^T\right]^T.
\end{equation}
Thus, the power of MUI terms is formulated as follows
\begin{align}\label{power int term irs}
    P_{MUI}^{IRS}&=\left|\left|\mb{e}\right|\right|^2
    \nonumber\\&=\mb{\Psi}_{I}^{\vect^H}\mb{A}\Psi_{I}^{\vect}-\mb{b}^{H}\mb{\Psi}_{I}^{\vect}-\mb{\Psi}_{I}^{\vect^H}\mb{b}+E_{s}^{\prime},
\end{align}
where
\begin{align}
    &\mb{A}=\bs{\Pi}^{H}\bs{\Pi},\\
    &\mb{b}=\bs{\Pi}^{H}\bs{\eta},\\
    &E_{s}^{\prime}=\left|\left|\bs{\eta}\right|\right|^2.
\end{align}
As a consequence, in order to optimize the IRS, the following optimization problem should be solved
\begin{equation}\label{main non convex opt for IRS}
\begin{array}{cl}
\min\limits_{\Psi_I}& \mb{\Psi}_{I}^{\vect^H}\mb{A}\mb{\Psi}_{I}^{\vect}-\mb{b}^{H}\mb{\Psi}_{I}^{\vect}-\mb{\Psi}_{I}^{\vect^H}\mb{b}  \\ \\
\st
& \diag{(\mb{\Psi}_{I}^{\vect}\mb{\Psi}_{I}^{\vect^H})}=\mb{1}_{M\times 1} \\ \\
&\psi_{j}\in \mathbb{D}_{N_{IRS}},\quad j=1, \dots, M.
\end{array}
\end{equation}
By defining $\mb{\Psi}_I^{\prime}=[\mb{\Psi}_{I}^{\vect^T},1]^T$ and
$\mb{\Omega}= \left[
    \begin{array}{cc}
        \mb{A} &  -\mb{b} \\
        -\mb{b}^H
          & 0
    \end{array}\right]$  the equivalent optimization problem can be written as
    \begin{equation}\label{equi2 non convex opt for IRS}
\begin{array}{cl}
\min\limits_{\Psi^{\prime}}& \tr\left(\mb{\Omega}\mb{\Psi}^{\prime}\right)  \\ \\
\st
& \diag{(\mb{\Psi}^{\prime})}=\mb{1}_{(M+1)\times 1} \\ \\
&\rank(\mb{\Psi}^{\prime})=1,\\ \\
&\mb{\Psi}^{\prime}\succeq 0,\\
&\psi_{j}\in \mathbb{D}_{N_{IRS}},\quad j=1, \dots, M,
\end{array}
\end{equation}
where $\mb{\Psi}^{\prime}=\mb{\Psi}_I^{\prime}\mb{\Psi}_I^{\prime\ H}$.
If we relax the optimization problem in (\ref{equi2 non convex opt for IRS}) by ignoring the rank-one constraint, the following  problem is obtained
\begin{equation}\label{Relaxed equi2 convex opt for IRS}
\begin{array}{cl}
\min\limits_{\Psi^{\prime}}& \tr\left(\mb{\Omega}\mb{\Psi}^{\prime}\right)  \\ \\
\st
& \diag{(\mb{\Psi}^{\prime})}=\mb{1}_{(M+1)\times 1} \\ \\
&\mb{\Psi}^{\prime}\succeq 0.
\end{array}
\end{equation}
As before, if $\mb{\Psi}^{\prime}$ is a rank-one matrix, $\mb{\Psi}_I^{\prime}$ and $\mb{\Psi}_I^{\vect}$ are directly recovered from $\mb{\Psi}^{\prime}$ and mapped into the discrete phase space. However, if the rank of this matrix is  $r>1$, the following steps should be conducted:

 \begin{enumerate}
      \item Solve the SDP relaxation problem in (\ref{Relaxed equi2 convex opt for IRS})  to obtain an optimal solution $\mb{\Psi}^{\prime \ opt}$. Since
$\mb{\Psi}^{\prime \ opt}$ is a positive semi-definite matrix,  the Cholesky decomposition could be utilized as
$\mb{\Psi}^{\prime \ opt}=\mb{R}\mb{R}^{H}$. 
      \item Generate a random vector $\mb{v}$ with the distribution of $\mb{v}\sim\mathcal{CN}(\mb{0},\mb{I}_{M+1})$.
      \item For $s=1, 2,\dots, M+1$, let $\hat{\psi}_{s} = g ({\mb{r}}_{s}\mb{v})$, where ${\mb{r}}_{s}$ denotes the $s$-th row of $\mb{R}$ and the function $g (.)$ is defined as follows
  \end{enumerate}
  \begin{align}
          &g(x)=\nonumber\\
          &\begin{cases}
          1, \ \quad\quad\quad\quad\quad \text{arg}(x)\in [  \frac{-\pi}{N_{IRS}}
          ,\frac{\pi}{N_{IRS}})\\ e^{\frac{\jj2\pi}{N_{IRS}}}, \ \ \quad\quad\ \text{arg}(x)\in [  \frac{\pi}{N_{IRS}}
          ,\frac{3\pi}{N_{IRS}})\\ \vdots \\ e^{\frac{\jj2\pi (N_{IRS}-1)}{N_{IRS}}}, \ \text{arg}(x)\in [  \frac{(2N_{IRS}-3)\pi}{N_{IRS}}
          ,\frac{(2N_{IRS}-1)\pi}{N_{IRS}}).
          \end{cases}
      \end{align}
Note that $\hat{\psi}_{s}\in \mathbb{D}_{N_{IRS}},  \  \text{for} \  s=1, \dots, M$ satisfies the constraint of the optimization problem in (\ref{main non convex opt for IRS}). In this case,  $\mb{\hat{\Psi}}_I=\diag([\hat{\psi}_{1}, \dots , \hat{\psi}_{M}])$ would be the final solution. \textcolor{\revcolor}{As before, repeating the randomization stage improves the accuracy of the solution. Also,}  this algorithm leads to an $\frac{(N_{IRS}\sin(\frac{\pi}{N_{IRS}}))^2}{2N_{IRS}}$-approximation error.

\section{Interference  Analysis}\label{interference analisys}
In this section, the performance of the IRS is investigated in terms of imposed MUI. To this end, the interference term in (\ref{Int term IRS}) is considered. For the sake of inducing the minimum of  MUI, the following term should be minimized
\begin{equation}\label{power of MUI2}
    \left|\left|\mb{e}^{[j]}\right|\right|^2=\left|\left|\bs{\Pi}^{[j]}\mb{\Psi}_{I}^{\vect}-\bs{\eta}^{[j]}\right|\right|^2.
\end{equation}
Note that if the number of users ($K$) is smaller than the number of IRS antennas ($M$), the  equation in (\ref{Int term IRS}) would be under-determined. As a consequence, if $\bs{\Pi}^{[j]}$ is a full-rank matrix, an interference-free precoder is achieved by
\begin{equation}
    \mb{\Psi}_{I,\text{opt}}^{\vect}=\mu^{[j]}\bs{\Pi}^{[j] \ H}\left(\bs{\Pi}^{[j]}\bs{\Pi}^{[j] \ H}\right)^{-1}\bs{\eta}^{[j]},
\end{equation}
in which the scaling factor $\mu{[j]}$ is set to fulfill the unit-power
constraint as follows
\begin{equation}
    {\mu^{[j]}}^2=\frac{1}{\tr\left(\bs{\Pi}^{[j] \ H}\bs{\Pi}^{[j]}\right)^{-1}}.
\end{equation}
Nonetheless, in practical scenarios, the matrix $\bs{\Pi}^{[j]}$ may not be necessarily full-rank. As a result, the interference term in (\ref{power of MUI2}) would be non-zero and consequently  the corresponding precoder undergoes  inevitable MUI.

Now, we aim to investigate the impact of the presence of the IRS from the achievable rate point of view. To this end, the power of MUI for the $i$-th user in the $j$-th cell is calculated as follows
\begin{align}\label{per user irs power}
    P_{MUI}^{IRS \ [i,j]}&=\left|\left|\mb{e}^{[i,j]}\right|\right|^2
    \nonumber\\&=\mb{\Psi}_{I}^{\vect^H}\bs{\Pi}^{[i,j] \ H}\bs{\Pi}^{[i,j]}\Psi_{I}^{\vect}\nonumber\\
    &-\mb{b}^{[i,j] \ H}\mb{\Psi}_{I}^{\vect}-\mb{\Psi}_{I}^{\vect^H}\mb{b}^{[i,j]}+\left|\eta^{[i,j]}\right|^2,
\end{align}
where $\bs{\Pi}^{[i,j]}$ and $\eta^{[i,j]}$ are the $i$-th row of $\bs{\Pi}^{[j]}$ and the $i$-th entry of $\bs{\eta}^{[j]}$, respectively and $\mb{b}^{[i,j]}=\eta^{[i,j]}\bs{\Pi}^{[i,j] \ H}$. For the sake of brevity, the equation in (\ref{per user irs power}) is rewritten by
\begin{equation}\label{equi IRS MUI power}
    P_{MUI}^{IRS \ [i,j]}=J^{[i,j]}(\Psi_{I}^{\vect})+P_{MUI}^{[i,j]},
\end{equation}
in which, $P_{MUI}^{[i,j]}=|\eta^{[i,j]}|^2$ is the power of MUI for the $i$-th user of the $j$-th cell in absence of the IRS and
\begin{align}\label{IRS power term}
    J^{[i,j]}(\Psi_{I}^{\vect})&=\mb{\Psi}_{I}^{\vect^H}\bs{\Pi}^{[i,j] \ H}\bs{\Pi}^{[i,j]}\Psi_{I}^{\vect}\nonumber\\
    &-\mb{b}^{[i,j] \ H}\mb{\Psi}_{I}^{\vect}-\mb{\Psi}_{I}^{\vect^H}\mb{b}^{[i,j]}.
\end{align}
It is worth noting that, the presence of the IRS would result in either a constructive impact or a destructive impact on the system performance based on the value of (\ref{IRS power term}). To be specific, since $P_{MUI}^{[i,j]}$ is a positive value, in order to achieve a constructive impact, $J^{[i,j]}(\Psi_{I}^{\vect})$ in (\ref{equi IRS MUI power}) should be negative. On the other hand, due to the positive value of $P_{MUI}^{IRS \ [i,j]}$, it can be concluded that $|J^{[i,j]}(\Psi_{I}^{\vect})|<P_{MUI}^{[i,j]}$.
According to the analysis conducted in \cite{Mohammed2013Per-AntennaSystems}, it can be shown that for an interference-dominant environment ($P_{MUI}^{IRS \ [i,j]}\gg \sigma_{w}^{2})$, the achievable rate for the $i$-th user in the $j$-cell is obtained as follows
\begin{align}\label{achievable rate}
    r_{IRS}^{[i,j]}&=\log_{2}\left(\frac{\left|s^{[i,j]}\right|^2}{\mathbb{E}_{\mb{g}}\left[P_{MUI}^{IRS \ [i,j]}\right]}\right)\nonumber\\
    &=\log_{2}\left(\frac{\left|s^{[i,j]}\right|^2}{\mathbb{E}_{\mb{g}}\left[J^{[i,j]}(\Psi_{I}^{\vect})\right]+\mathbb{E}_{\mb{g}}\left[P_{MUI}^{[i,j]}\right]}\right).
\end{align}
In order to derive a lower bound for the achievable rate, a limited impact for the IRS is considered, i.e, $|\mathbb{E}_{\mb{g}}\left[J^{[i,j]}(\Psi_{I}^{\vect})\right]|\ll \mathbb{E}_{\mb{g}}[P_{MUI}^{[i,j]}]$. In this case, assuming $\varepsilon=|\mathbb{E}_{\mb{g}}\left[J^{[i,j]}(\Psi_{I}^{\vect})\right]|/\mathbb{E}_{\mb{g}}[P_{MUI}^{[i,j]}]$ and utilizing the approximation $1/(1-\varepsilon)\approx1+\varepsilon, \ \varepsilon\ll 1$, the achievable rate in (\ref{achievable rate}) can be approximated by
\begin{align}
  r_{IRS}^{[i,j]}&\approx \log_{2}\left(\frac{\left|s^{[i,j]}\right|^2}{\mathbb{E}_{\mb{g}}\left[P_{MUI}^{[i,j]}\right]}\right)+\frac{\left|\mathbb{E}_{\mb{g}}\left[J^{[i,j]}(\Psi_{I}^{\vect})\right]\right|}{\mathbb{E}_{\mb{g}}\left[P_{MUI}^{[i,j]}\right]}\log_{2}(e)\nonumber\\
  &=r_{IRS-free}^{[i,j]}+\frac{\left|\mathbb{E}_{\mb{g}}\left[J^{[i,j]}(\Psi_{I}^{\vect})\right]\right|}{\mathbb{E}_{\mb{g}}\left[P_{MUI}^{[i,j]}\right]}\log_{2}(e),
\end{align}
where $r_{IRS-free}^{[i,j]}$ is the achievable rate for the $i$-th user of the $j$-th cell in the absence of the IRS. Thanks to the positive value of $\mathbb{E}_{\mb{g}}[P_{MUI}^{[i,j]}]$, it is obvious that $r_{IRS}^{[i,j]}>r_{IRS-free}^{[i,j]}$. Hence, it can be concluded that, in presence of the IRS, the lower bound of the achievable rate would be greater than the absolute achievable rate of the IRS-free scenario.

\section{Complexity Analysis}\label{Complexity Analysis}
This section is dedicated to analyzing the computational complexity of each scheme.
The proposed  trellis-based algorithms consist of  $N_{T}-T$ and $M-B$ stages, respectively. The optimization for \textcolor{\revcolor}{$\mb{x}_{j}$} consists of $(N_{BS})^{T}$ states with $N_{BS}$ branches entering each state. Similarly,  the optimization for $\Psi_{I}$ contains $(N_{IRS})^{B}$ states with $N_{IRS}$ branches entering each state. Hence, a total of $(N_T-T)(N_{BS})^{T+1}+(M-B)(N_{IRS})^{B+1}$ comparisons are needed. This number is negligible compared with an exhaustive search method with $(N_{BS})^{N_T}+(N_{IRS})^{M}$ comparisons, especially for large numbers of $N_T$ and $M$.
Note that, utilizing the interior point algorithm and assuming a solution accuracy $\epsilon>0$, the computational complexity of the SDR method is \textcolor{\inquirycolor}{of the order $\mathcal{O}(((N_T+1)^{3.5}+(M+1)^{3.5})\log(1/ \epsilon)))$} \cite{Helmberg1996} which is not scalable for cases with large dimensions.

To better understand the complexity issue, consider a scenario where $N_T=50$, $M=70$, and $T=B=2$. Based on these assumptions, we consider two cases: Case 1 \textcolor{\revcolor}{with} $N_{BS}=N_{IRS}=4$; \textcolor{\revcolor}{and} Case 2 \textcolor{\revcolor}{with} $N_{BS}=N_{IRS}=8$. The computational complexity of the various schemes is presented in Table  \ref{table 1}.
\begin{table}[ht]
\caption{Complexity of the Algorithms}
\centering
\begin{tabular}{c c c c}
\hline\hline
Case & Trellis-based scheme & SDR-based scheme & Exhaustive search \\ [0.5ex]
\hline
1 & $7.42*10^3$ & $3.96*10^{6}$ & $1.39*10^{42}$ \\ \hline
2 & $5.94*10^{4}$ & $3.96*10^{6}$ & $1.64*10^{63}$\\\hline
\end{tabular}
\label{table 1}
\end{table}

\section{Numerical Evaluations}\label{Numerical Evaluations}
In this section, the simulation results are \textcolor{\quirycolor}{performed} to evaluate the \textcolor{\quirycolor}{performance}  of  all of the proposed schemes. The performance metric is the average data rate per cell, and the results are obtained \textcolor{\revisioncolor}{by averaging over 100 independent channel generations.}  Unless otherwise stated, it is assumed that the system model consists of $L=5$ cells, each of which has a BS with $N_T=50$ antennas and $K=15$ single-antenna users. In addition, an IRS with $M=70$ discrete $3$-bit PSs is used \textcolor{\revisioncolor}{to assist the downlink transmission of BSs.}
The total transmit power of each BS is set to  $P_T=3 W$ and the noise power in all scenarios is $-30$dB. The desired symbol is modulated by the quadratic phase shift keying (QPSK), and the memory of the trellis structures in both the IRS and the BSs is assumed to be $T=B=2$. \textcolor{\revisioncolor}{Also, the desired symbols are assumed to be with unit power, i.e. $\big|s^{[i,j]}\big|^2=1$, $i=1,\dots,K$, $j=1,\dots,L$.}

\begin{figure}[t]
\begin{center}
   \includegraphics[scale=.64]{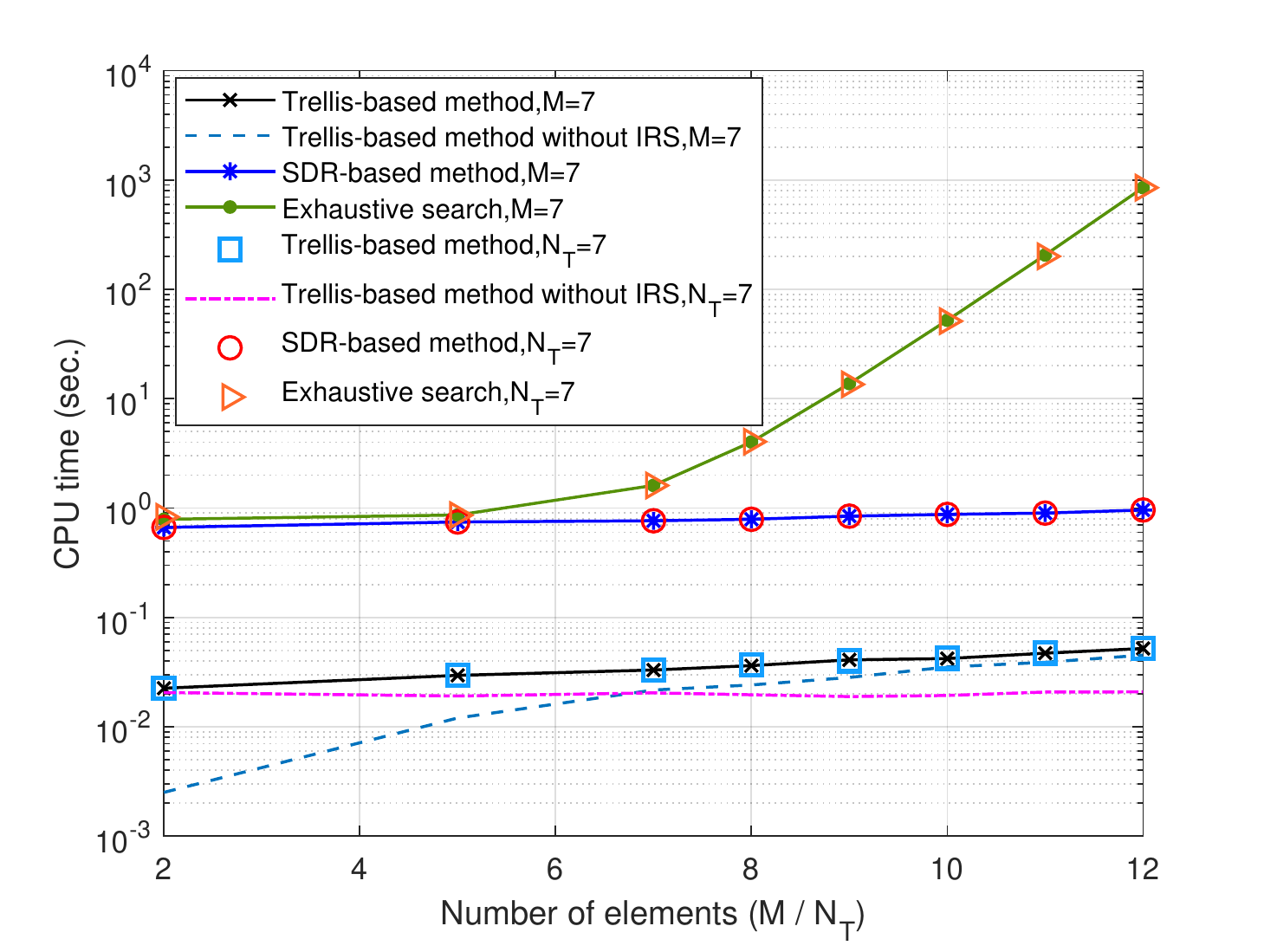}
       \caption{CPU time versus number of IRS phase shifters, in case of a single-cell system with $N_T=7$ BS antennas and $K=5$ single-antenna users.}
    \label{fig:Complexity}
\end{center}
\end{figure}

Fig. \ref{fig:Complexity} depicts the complexity of the presented algorithms for a simple single-cell scenario in which an IRS aids the downlink transmission of a BS with $N_T=7$ antennas to $K=5$ single-antenna users. It is assumed that the resolution of the PSs is 2 bits. The results are obtained by using a computer with a 1.60 GHz i5-8265U CPU and 8 GB RAM.  The dimension of the system is very small since for higher dimensions, the exhaustive search algorithm would take a long time to run. \textcolor{\revcolor}{In Fig. \ref{fig:Complexity},} the CPU time for performing the algorithms is shown based on the number of elements in either the BS or the IRS. If $N_T$ is variant, $M=7$ is constant and if $M$ is variant $N_T=7$ is constant. As described in this figure, as the number of elements grow\textcolor{\revcolor}{s}, the complexity of the exhaustive search algorithm increases dramatically, while the complexity of trellis-based and SDR-based algorithms rises with a much slower \textcolor{\revcolor}{speed}.  \textcolor{\revcolor}{Also}, the CPU time for the trellis-based algorithm is much lower than the SDR-based method. Moreover, it is depicted that adding one IRS into the system and using the trellis-based method for its beamforming operation does not add much to the computational complexity.

\begin{figure}[t]
\begin{center}
   \includegraphics[scale=.64]{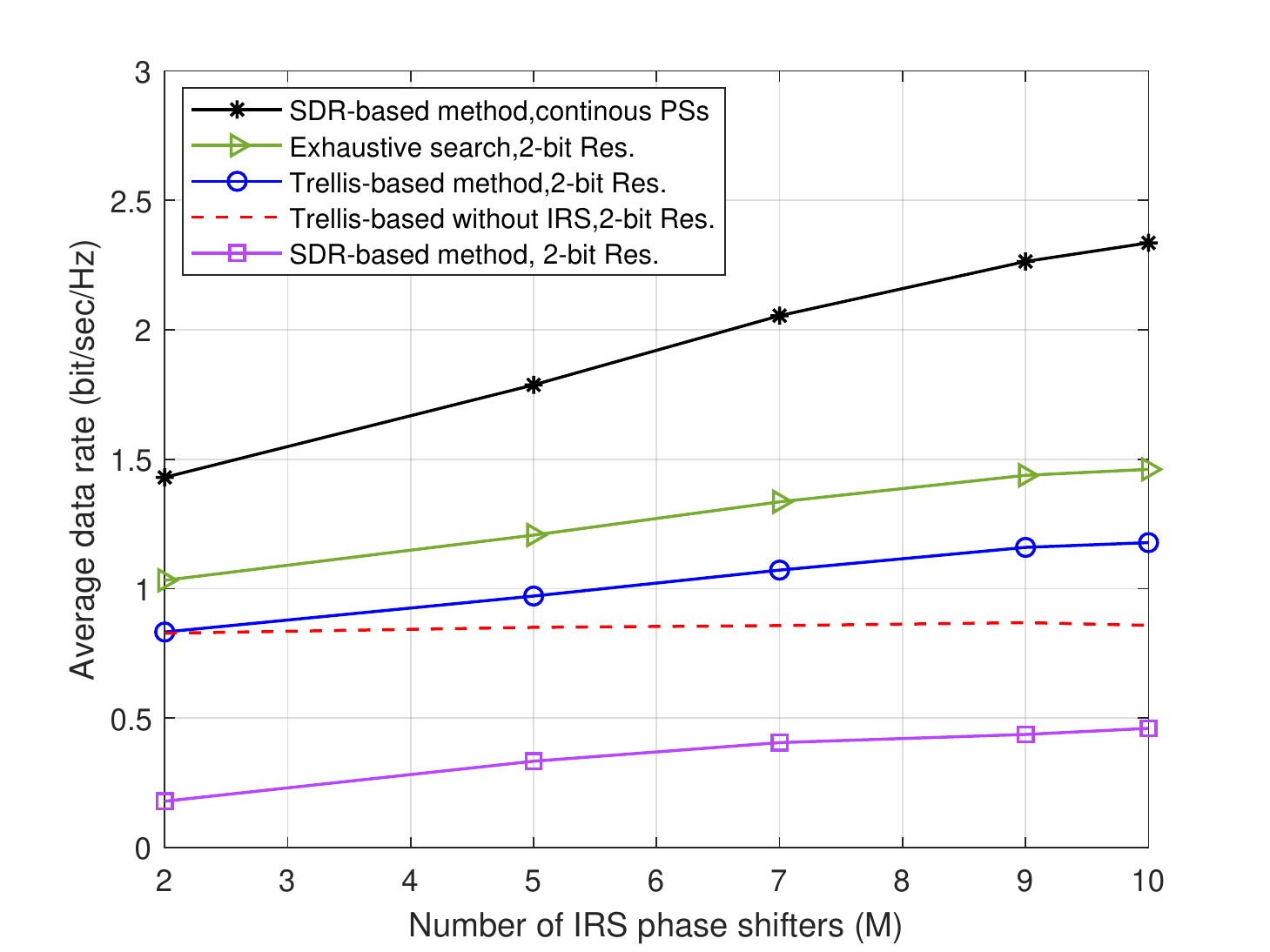}
       \caption{Average data rate versus the number of IRS reflectors, in case of a single cell system with a $N_T=7$ antenna BS and $K=10$ single-antenna users.}
    \label{fig:RATE_M_SIMPLE}
\end{center}
\end{figure}

Fig. \ref{fig:RATE_M_SIMPLE} depicts the average data rate versus the number of IRS PSs ($M$), while using different methods of joint IRS beamforming and BS precoding. In this figure, the system is single-cell with $N_T=7$ antennas at the BS and $K=10$ single-antenna users. The dimension of the system is small since we aim to compare the performance to the exhaustive search algorithm. The SDR-based method with infinite-resolution PSs is also presented to determine an upper bound for the proposed algorithms. The trellis-based method with 2-bit PSs is realized  once with an IRS and once without an IRS in the system, and it is shown that by increasing $M$, the average data rate is increased due to the provided degrees of freedom. As shown in this figure, the performance of the trellis-based method is close to that of the exhaustive search algorithm. It is also demonstrated that the SDR-based method does not perform well when low-resolution PSs are utilized. Hence, it can be concluded that the trellis-based method is \textcolor{\revcolor}{concurrently} the best solution for systems with discrete PSs, in terms of both performance and complexity.

\begin{figure}[t]
\begin{center}
   \includegraphics[scale=.64]{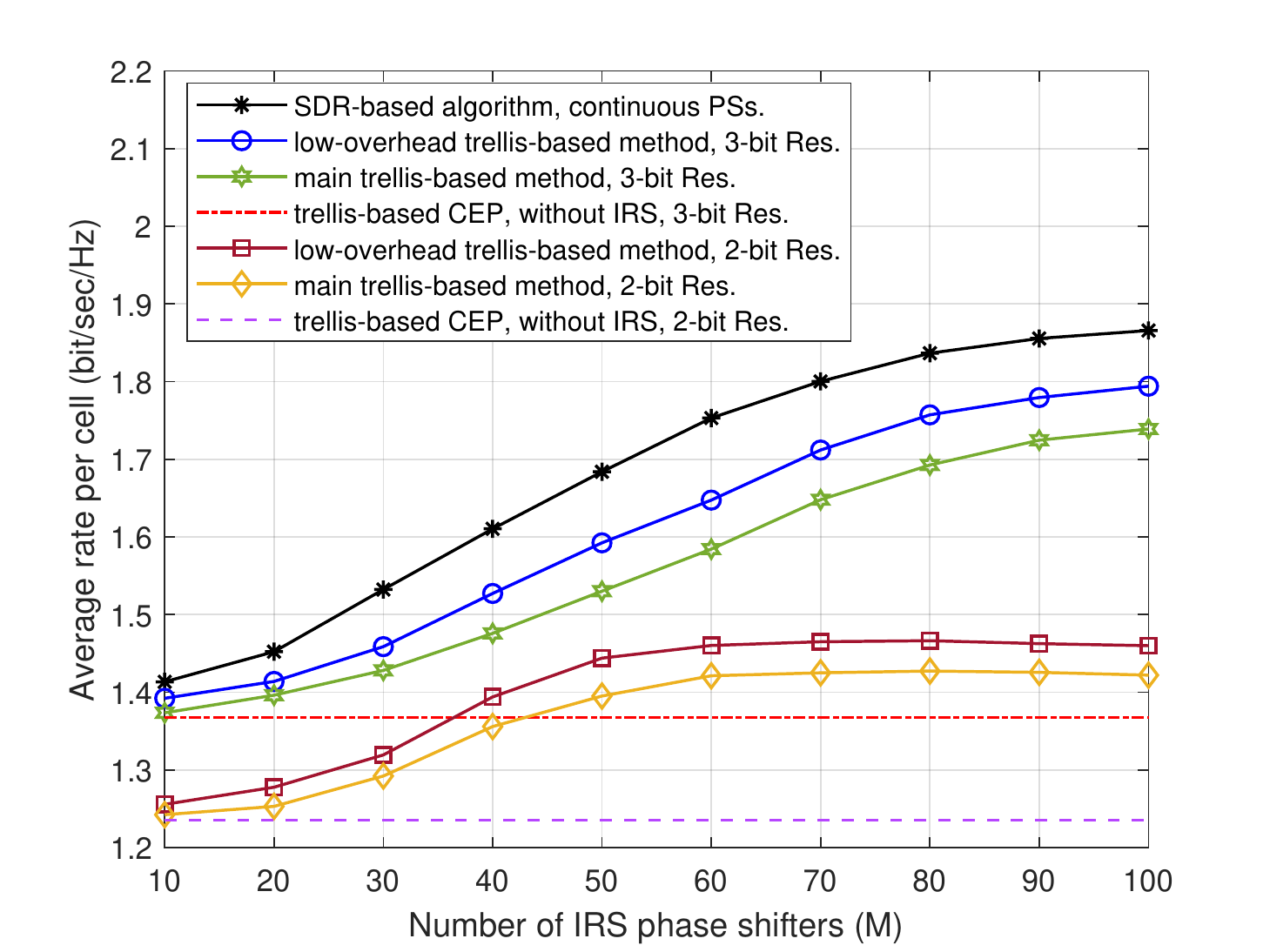}
       \caption{Average data rate per cell versus number of IRS reflectors, a comparison between different schemes with different resolutions of PSs.}
    \label{fig:R-M-Res4}
\end{center}
\end{figure}

In Fig. \ref{fig:R-M-Res4}, a multi-cell IRS-aided large-scale MIMO system is considered where an IRS aids the downlink transmission of $L=5$ BSs and mitigates the inter-cell interference. It is assumed that each cell contains $K=15$ single-antenna users and a BS with $N_T=50$ CE antennas. The resolution of IRS and BS PSs is once set to 2 and once again set to 3 bits in the case of using trellis based approaches. In the SDR-based method, continuous PSs are used to provide an upper bound for the presented schemes.
In this figure, the average data rate for four scenarios is plotted versus the number of IRS reflectors. As shown in this figure, in the case when there is no IRS in the system, the average data rate is  low, while \textcolor{\inquirycolor}{by deploying an IRS into our $5$-cell system model}, the performance is greatly enhanced, and as the number of PSs increases, \textcolor{\revisioncolor}{due to availability of more degrees of freedom}, the data rate rises. \textcolor{\quirycolor}{For the low-overhead trellis-based algorithm, the performance is higher than that of the main trellis-based scheme, which is due to the minimization of inter-cell interference terms by each BS.}
\textcolor{\revisioncolor}{It is interesting to observe that, by comparing the results of $2$-bit PSs to that of $3$-bit PSs, it can be found that adding only $1$ bit to the resolution of PSs can significantly enhance the system performance.
The SDR-based algorithm, on the other hand, is presented \textcolor{\revisioncolor}{as a  benchmark method to be compared}. This method, which can only provide acceptable solutions in the case of  high resolution PSs, can only achieve slightly higher data rate than the trellis-based schemes at the expense of high implementation cost.}

\begin{figure}[t]
\begin{center}
   \includegraphics[scale=.64]{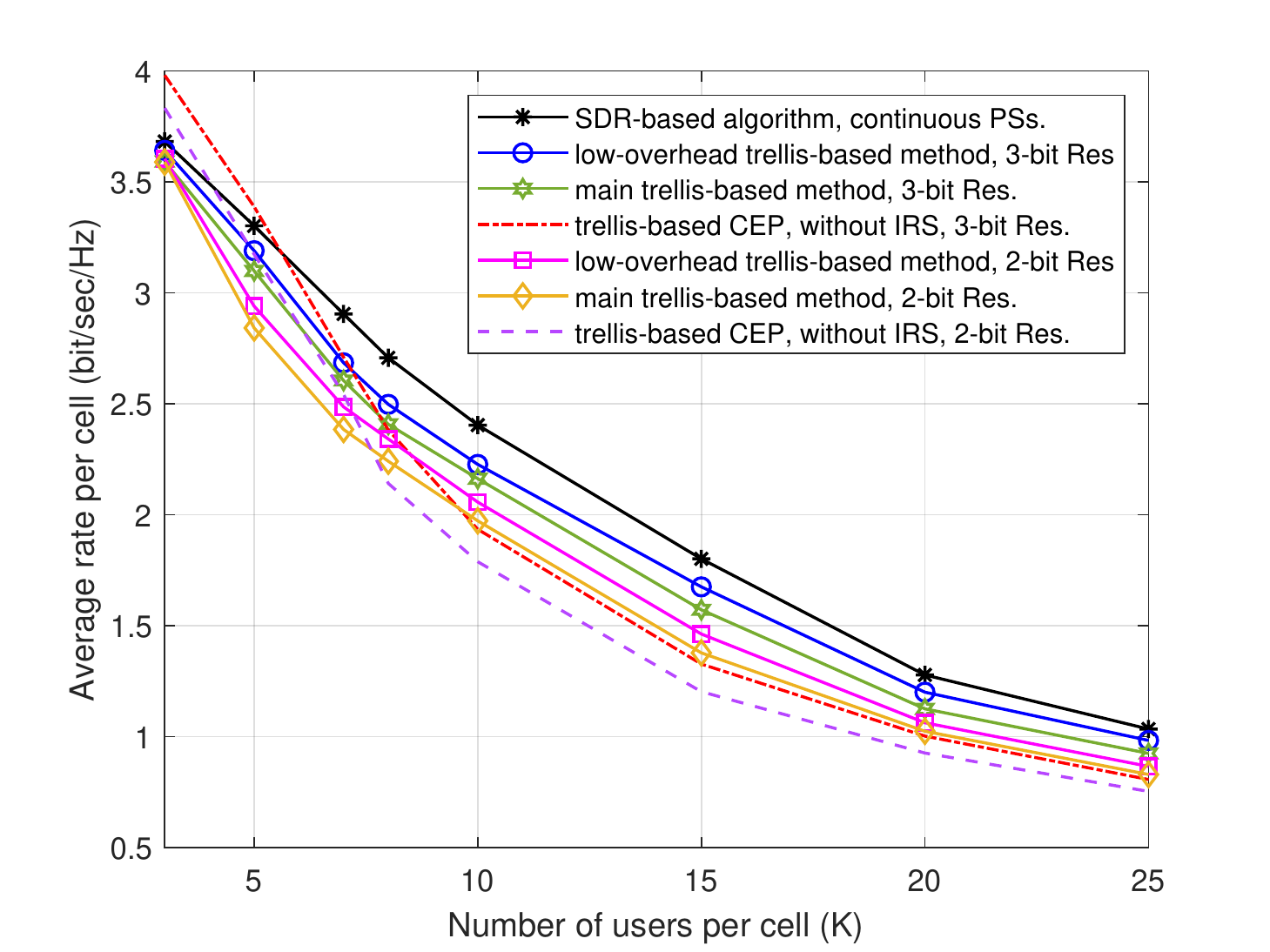}
       \caption{Average data rate per cell versus number of users per cell,  a comparison between different schemes with different resolutions of PSs.}
    \label{fig:R-K}
\end{center}
\end{figure}

Fig. \ref{fig:R-K} investigates the effect of number of users on the system performance. Here, it is assumed that there are $L=5$ cells in the system, each containing a BS with $N_T=50$ antennas and different numbers of users. An IRS equipped with $M=70$ phase shifters, serves all cells. The SDR-based approach, as before, uses continuous PSs in the IRS and BSs, \textcolor{\revisioncolor}{while other methods are realized with $2$- and $3$-bit PSs}. As shown in this figure, by increasing the number of users in the system, the performance is degraded, since the available degrees of freedom are reduced. The use of IRS in the system is appealing for larger number of users in the system as the data rate for $K>7$ is higher, \textcolor{\revcolor}{in case the multi-cell system is enabled with the IRS.} For lower values of $K$, as discussed in Section \ref{interference analisys}, the systems does not suffer much MUI; \textcolor{\revcolor}{hence, despite the degrees of freedom caused by the IRS, the imposed interference component would be dominant.} In addition, as depicted in this figure, by increasing the resolution of the PSs in IRS and BSs, the average data rate is much enhanced.

\begin{figure}[t]
\begin{center}
   \includegraphics[scale=.64]{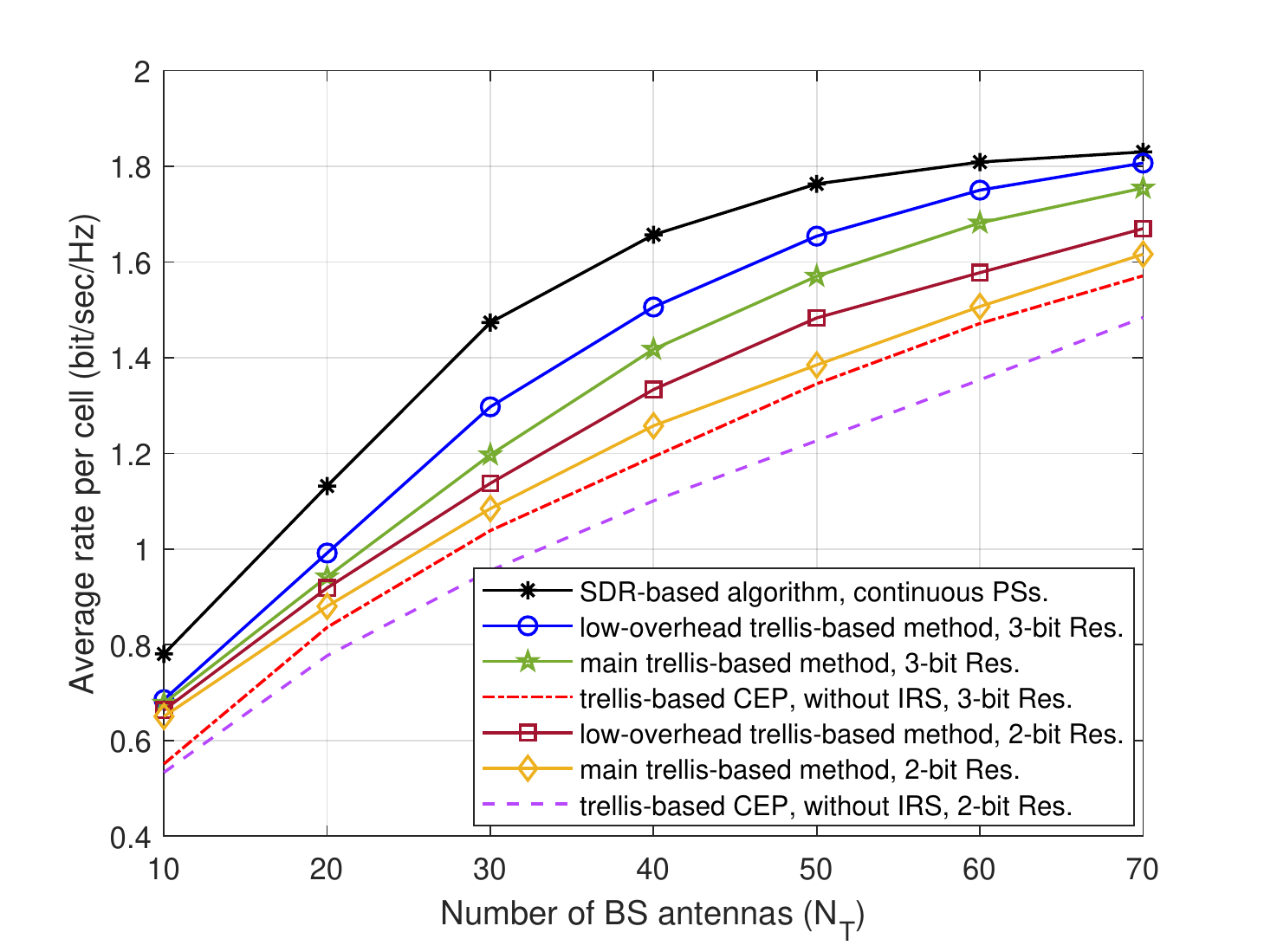}
       \caption{Average data rate per cell versus number of BS antennas,  a comparison between different schemes with different resolutions of PSs.}
    \label{fig:R-NT}
\end{center}
\end{figure}

In Fig. \ref{fig:R-NT}, the effect of the number of BS antennas is studied on the average data rate. It is assumed that there are $L=5$ cells in the system, each containing a BS and $K=15$ users, and an IRS with $M=70$ PSs serves all cells. The first thing to notice, is that the performance is enhanced as we increase $N_T$, which is obviously a result of the provided degrees of freedom by massive MIMO.  Secondly, the effect of IRS is justified especially in a lower number of BS antennas, where \textcolor{\revisioncolor}{MUI} is dominant. Last but not the least, as it is demonstrated, the difference between \textcolor{\revcolor}{the achieved average data rate of the SDR-based scheme and that of the trellis-based algorithms decreases, in higher values of $N_T$.} \textcolor{\inquirycolor}{This is due to the fact that the SDR-based method relies upon finding the rank-one solution, and if the solution is not rank-one, then, through randomization, it achieves a result \textcolor{\revcolor}{that is suboptimal}.ffffff Based on our simulations, when the number of BS antennas is high, the probability of obtaining rank-1 solution decreases.} Hence, when $N_T=60$ or $N_T=70$, the SDR-based scheme has higher probability to use  the randomization method. It can be concluded that, the proposed trellis-based schemes, not only \textcolor{\revisioncolor}{benefit from} lower consumed power and computational and hardware complexity, but also \textcolor{\revcolor}{result in more convenient results in higher number of BS antennas, specially when $3$-bit PSs are utilized in the BSs and the IRS.}

\section{Conclusion}\label{Conclusion}
In this paper, a large-scale MIMO system is considered where a discrete-phase IRS aids the downlink transmission. Here, for the first time, passive precoding methods are used for both BSs and IRS.
The purpose is to minimize the sum power of MUI by jointly optimizing the IRS beamforming and the BS precoding vectors. As the BSs use CEP, the problem is to select the best phase shifts for the BSs as well as the IRS to reach the minimum MUI. To this end, at first a  trellis-based joint IRS and BS precoding design is introduced, where the BS precoding operation in each cell is performed individually. Afterwards, by applying stochastic optimization, a low-overhead trellis-based optimization technique is presented, in which by minimizing inter-cell and intra-cell interference terms, a higher performance is achieved. Finally, for the sake of having a comparison benchmark, the SDR is applied to solve the problem. Moreover, an interference and complexity analysis is presented for all of the schemes, and the algorithms are compared to one another in terms of their achieved average data rate and CPU time.

\bibliographystyle{IEEEtran}
\bibliography{Mendeley,Ref1}
\end{document}